\documentclass[lettersize,journal]{IEEEtran}
\usepackage{amsmath,amsfonts}
\usepackage{array}
\usepackage[caption=false,font=normalsize,labelfont=sf,textfont=sf]{subfig}
\usepackage{textcomp}
\usepackage{stfloats}
\usepackage{url}
\usepackage{verbatim}
\usepackage{graphicx}
\usepackage{cite}
\usepackage{mathtools}
\usepackage{algorithm,algpseudocode}
\usepackage{amsmath}
\usepackage{booktabs}
\DeclareMathOperator*{\argmax}{arg\,max}
\DeclareMathOperator*{\argmin}{arg\,min}

\usepackage[short]{optidef}
\usepackage{bbm}
\usepackage[nolist,nohyperlinks]{acronym}
\usepackage{multirow, makecell, boldline}

\usepackage{xcolor}

\begin{document}

\title{Uncrewed Vehicles in 6G Networks: A Unifying Treatment of Problems, Formulations, and Tools}

\author{Winston Hurst, Spilios Evmorfos, Athina Petropulu, Yasamin Mostofi

\thanks{Winston Hurst and Yasamin Mostofi are with the Department of Electrical and Computer Engineering, University of California, Santa Barbara, CA 93117 USA (email: \{winstonhurst, ymostofi\}@ece.ucsb.edu).\\
Spilios Evmorfos and Athina P. Petropulu are with the Department of
Electrical and Computer Engineering, Rutgers, The State University of New
Jersey, Piscataway, NJ 08854 USA (email: se386@scarletmail.rutgers.edu;
athinap@soe.rutgers.edu).
}
\thanks{This work was supported in part by NSF RI award 2008449, ONR award N00014-23-1-2715, and ARO Grants W911NF2110071 and W911NF2320103.}}

\IEEEpubid{\begin{minipage}{\textwidth}\centering
    {\vspace{0.75in}\footnotesize \copyright 2025 IEEE.  Personal use of this material is permitted.  Permission from IEEE must be obtained for all other uses, in any current or future media, including reprinting/republishing this material for advertising or promotional purposes, creating new collective works, for resale or redistribution to servers or lists, or reuse of any copyrighted component of this work in other works.}
\end{minipage}}
\maketitle

\begin{abstract}
Uncrewed Vehicles (UVs) functioning as autonomous agents are anticipated to play a crucial role in the 6th Generation of wireless networks. Their seamless integration, cost-effectiveness, and the additional controllability through motion planning make them an attractive deployment option for a wide range of applications, both as assets in the network (e.g., mobile base stations) and as consumers of network services (e.g., autonomous delivery systems). However, despite their potential, the convergence of UVs and wireless systems brings forth numerous challenges that require attention from both academia and industry.  This paper then aims to offer a comprehensive overview encompassing the transformative possibilities as well as the significant challenges associated with UV-assisted next-generation wireless communications. Considering the diverse landscape of possible application scenarios, problem formulations, and mathematical tools related to UV-assisted wireless systems, the underlying core theme of this paper is the unification of the problem space, providing a structured framework to understand the use cases, problem formulations, and necessary mathematical tools.  Overall, the paper sets forth a clear understanding of how uncrewed vehicles can be integrated in the 6G ecosystem, paving the way towards harnessing the full potential at this intersection. 
\end{abstract}


\section{Introduction}\label{sec:intro}

6G systems are envisioned to make significant progress in the advancement of wireless communication technology, as they promise unparalleled capabilities that far surpass those of 5G systems \cite{OJCS2021Jiang, Network2020Saad, CommsMag2020Giordani}. In addition to improving reliability, latency, and capacity by orders of magnitude, 6G will introduce new dimensions to system services, including those related to sensing, artificial intelligence (AI), and positioning, all while responding to societal challenges such as climate change and inequity through an emphasis on sustainability and closing the digital divide \cite{IMT-2030}.

\begin{table}[]
    \centering
        \caption{List of Key Abbreviations.}
    \def\arraystretch{1.3}
    \begin{tabular}{||c|c||}
        \hline
        3GPP & Third Generation Partnership Project\\
        \hline
        5G & Fifth generation \\
        \hline
        6G & Sixth generation \\  
        \hline
        BCD & Block Coordinate Descent\\
        \hline
        BS & Base Station \\ 
        \hline
        DRL & Deep Reinforcement Learning\\
        \hline
        HAPS & High-altitude Platform Station  \\
        \hline
        ITU & International Telecoms Union\\
        \hline
        LOS & Line-of-Sight\\
        \hline
        MIMO & Multiple-input, multiple output\\
        \hline
        mmWave & Millimeter Wave  \\
        \hline
        MIP & Mixed-integer Program\\
        \hline
        MILP & Mixed-integer Linear Program\\
        \hline
        MINLP & Mixed-integer Nonlinear Program\\
        \hline
        NLOS & Non-LOS\\
        \hline
        NOMA & Non-Orthogonal Multiple Access\\
        \hline
        OFDMA & Orthogonal Frequency Division Multiple Access\\
        \hline
        OMA & Orthogonal Multiple Access\\
        \hline
        OTFS & Orthogonal Time Frequency Space\\
        \hline
        PD-NOMA & Power-Domain NOMA\\
        \hline
        PRB & Physical Resource Block\\
        \hline
        QoS & Quality of Service\\
        \hline
        RL & Reinforcement Learning\\
        \hline
        RSMA & Rate-Splitting Multiple Access\\
        \hline
        SCA & Successive Convex Approximation\\
        \hline
        SDMA & Space-division Multiple Access\\
        \hline
        SOCP & Second-order Cone Program\\
        \hline
        TBS & Terrestrial Base Station\\
        \hline
        TDMA & Time-division Multiple Access\\
        \hline
        THz & Terahertz\\
        \hline
        UAV & Unmanned Aerial Vehicle  \\
        \hline
        UV & Unmanned Vehicle  \\
        \hline
        UV-BS & UV acting as a BS\\
        \hline
        UGV & Unmanned Ground Vehicle  \\ 
        \hline
    \end{tabular}
    \label{tab:abbreviations}
    \vspace{-0.2in}
\end{table}

There are a number of inter-related challenges to overcome before realizing the full 6G vision. Some of these are posed directly by the stated network capabilities. How can ubiquitous coverage be achieved in areas where establishing network infrastructure is difficult, such as in maritime environments or rural areas \cite{WC2022Chaoub}? Others arise as we consider use cases. For instance, how can throughput, reliability, and latency improvement be guaranteed in dense Internet of Everything (IoE) networks \cite{Network2020Saad}, or for extended reality applications \cite{samsung_6g}? Still more challenges emerge as we consider issues associated with specific technologies, such as the acute attenuation and penetration loss of terahertz (THz) communication \cite{Access2019Rappaport}.

Amidst these challenges, one transformative player emerges in the 6G landscape: \textit{uncrewed vehicles}\footnote{We use the terms ``uncrewed vehicles" and ``unmanned vehicles" interchangeably throughout the manuscript}. Autonomous, uncrewed vehicles (UVs) (also called unmanned vehicles) are envisioned to have a crucial impact in shaping the future of 6G cellular systems, enabling them to overcome the limitations of conventional communication paradigms. They will facilitate novel use cases, enhance network efficiency, improve system intelligence and resilience, and support the high demands of truly immersive applications. 

In this paper, we critically explore the dynamic intersection of unmanned vehicles and 6G cellular systems, providing a comprehensive and methodical survey that encompasses the underlying challenges, associated problem formulations, and essential mathematical tools. \textbf{At the core of this research endeavor lies a central theme — the unification of the problem space}. When considering the integration of unmanned vehicles with 6G systems, researchers encounter many different scenarios, myriad problem formulations, and an array of mathematical tools. This makes it quite challenging to comprehend the full spectrum of possibilities in this area and harness them effectively. To address this, this paper takes a fresh perspective by offering a unifying approach. Specifically, it provides a structured framework, with an emphasis on unifying the scenarios, underlying problem formulations, and necessary mathematical tools, to enable successful navigation of this complex domain. By doing so, we equip researchers, engineers, and stakeholders with a clear understanding of how to integrate unmanned vehicles in the 6G ecosystem, which can facilitate the development of innovative solutions that effectively utilize the opportunities presented at this intersection.

We next provide a brief summary of the envisioned capabilities and supported applications that 6G can provide, followed by an overview of various types of unmanned vehicles relevant to the 6G ecosystem. This then paves the way for our methodical characterization of the roles of UVs in the 6G landscape. We end the section with a summary of our contributions.

\vspace{-0.1in}
\subsection{6G: Use Cases and Enabling Technologies}
Recent years have seen much discussion and speculation about beyond-5G communication networks, their capabilities, enabling technologies, and supported applications. In June 2023, the International Telecoms Union (ITU) published the IMT-2030 Framework~\cite{IMT-2030-Framework}, which lays out an official road map for 6G standardization and brings a level of concreteness to the discussion. Core use cases, performance capabilities that support these use cases, and technologies that, in turn, enable the cited performance, are briefly introduced below.

\begin{figure}
    \centering
\includegraphics[width=3.4in]{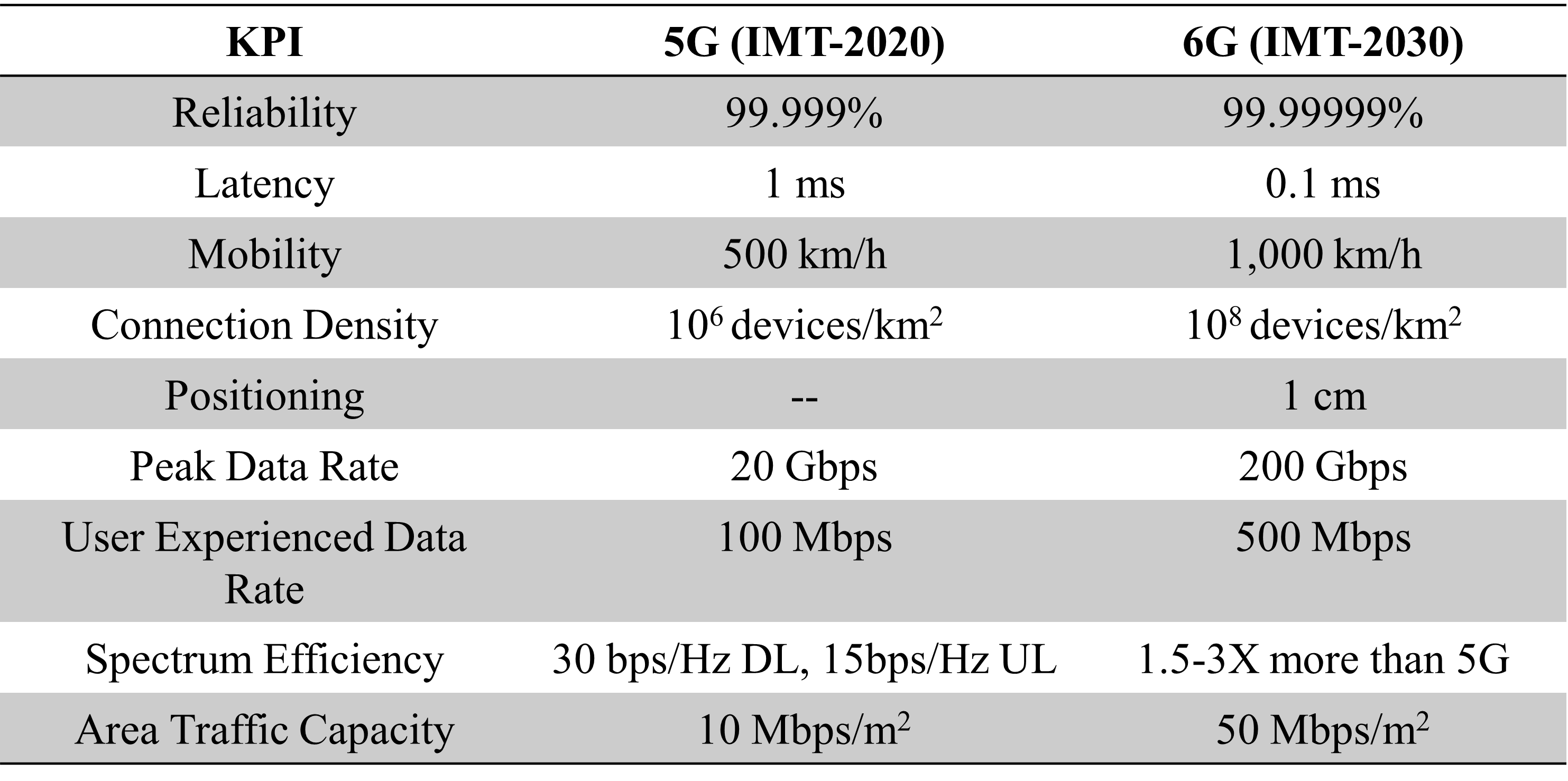}
\vspace{-0.1in}
    \caption{Key performance indicators (KPI) for 5G \cite{IMT-2020} and 6G \cite{IMT-2030}.}
    \label{fig:kpi_table}
    \vspace{-0.1in}
\end{figure}

\subsubsection{Use Cases and Applications}
Of the six core usage scenarios envisioned for 6G, three constitute extensions of 5G. Specifically, \textit{immersive communication}, \textit{massive communication}, and \textit
{Hyper Reliable and Low-Latency Communication} (HRLLC) extend enhanced mobile broadband (eMBB) for AR/VR applications \cite{samsung_6g}, enhance massive machine-type communication (mMTC), and improve on Ultra Reliable and Low-Latency Communication (URLLC), respectively. Three other core use cases are new in 6G. \textit{Integrated sensing and communication} (ISAC) will allow communication signals to also provide information about the physical environment, a feature useful for, \textit{e.g.},  autonomous vehicles navigating through dynamic urban areas \cite{JSAC2022Liu}. \textit{Ubiquitous connectivity} will close the digital divide, providing cellular and internet access to underserved areas. Finally, the 6G network will empower \textit{integrated artificial intelligence and communication}, facilitating intelligent, data-driven decision-making for many applications, from smart cities \cite{TIR2023Singh} to self-driving cars \cite{CommsMag2019Letaief}. 

\subsubsection{Enabling the 6G Vision}
Realizing the 6G vision will depend on underlying technical improvements to the network. Fig.~\ref{fig:kpi_table} illustrates several key performance indicators (KPI) and calls out how these improve over current 5G metrics. To make this enhancement possible, a number of important emerging technologies will be employed, as discussed next.

\begin{figure}
    \centering
    \includegraphics[width=3.4in]{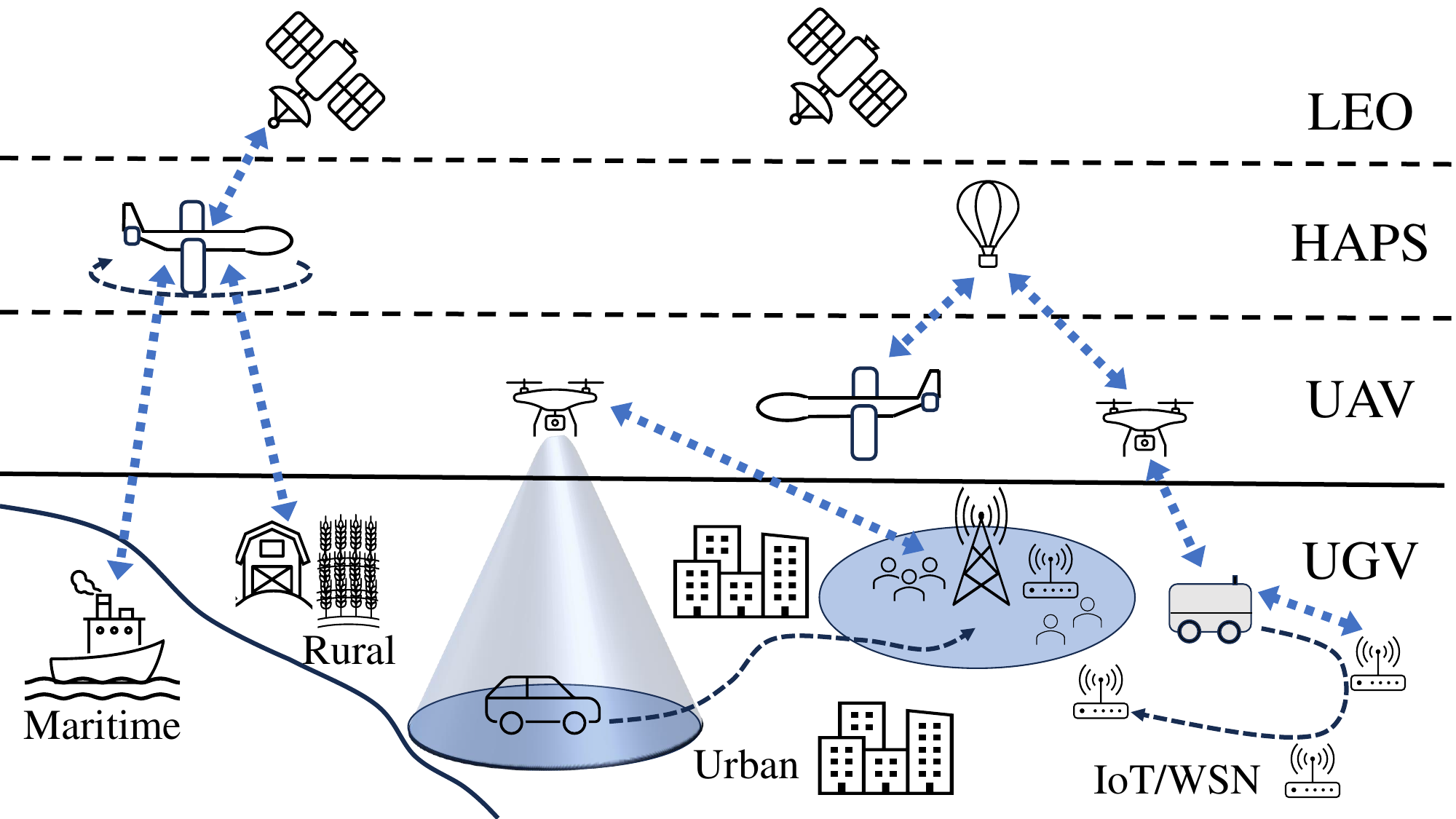}
    \vspace{-0.05in}
    \caption{A high-level view of the various roles UVs play in the 6G ecosystem.}
    \label{fig:ecosystem}
    \vspace{-0.2in}
\end{figure}

6G will see the maturation of mmWave communications and the introduction of sub-THz and THz communications. These bands raise new challenges, requiring advances in hardware design, channel modeling, and signal processing~\cite{Access2019Rappaport}. Extreme Multiple-Input-Multiple-Output (eMIMO)~\cite{IMT-2030} and reconfigurable intelligent surfaces (RIS)\cite{MobiCom2023Pallaprolu, SPM2022Bjornson,renzo2019smart} will permit greater directivity in transmissions, thereby overcoming fast attenuation and acute penetration loss experienced at higher frequencies~\cite{SF6GN2022DiRenzo}, while also enabling spatial multiplexing~\cite{FNWF2022Sun}. Other emerging multi-access methods, such as Power-Domain Non-Orthogonal Multiple Access (PD-NOMA) \cite{JNCA2022Ghafoor} and Rate-Splitting Multiple Access (RSMA) \cite{10038476}, promise to extract additional capacity from limited communication resources (time, frequency, and space) through intelligent signal processing and interference management. Finally, AI can enhance almost all aspects of the 6G ecosystem~\cite{SF6GN2022Stanczak}, including traffic prediction, channel learning, and resource management.

Particularly relevant for the discussion of UVs in 6G networks, the realization of the 6G vision depends on \textit{diverse network topographies}, as seen in Fig.~\ref{fig:ecosystem}. For example, Space-Air-Ground Integrated Networks (SAGIN) and Non-Terrestrial-Networks (NTNs) will be crucial in providing connectivity in scenarios where terrestrial-based cell systems fall short. In this context, UVs, especially Unmanned Aerial Vehicles (UAVs) and High-altitude Aerial Platform Stations (HAPS), will play a key role in realizing the 6G vision. We next discuss different types of UVs in more detail, before providing a more thorough discussion of their place in 6G and beyond.  

\vspace{-0.15in}
\subsection{Unmanned Vehicles}
Advances in robotics \cite{SOORI202354}, vision systems \cite{compVision2020}, and machine learning \cite{tengteng2021IJARS} of the past several years have led to increased excitement around the possibility of fully autonomous UVs, with envisioned use cases spanning deployment in air, on land, and at sea. In this paper, we focus on UAVs (including HAPS) and unmanned ground vehicles (UGVs).

UVs come in many shapes and sizes. For UGVs, wheeled vehicles, including car-like designs, remain a popular and important paradigm, but continuous track (treads) and legged robots also have advantages in certain environments. On the other hand, aerial vehicle designs generally follow either a fixed-wing or rotary-wing paradigm, though vertical takeoff and landing (VTOL) vehicles have garnered much interest in the past several years, particularly in the context of urban air mobility. Regardless of the type of UV, there is growing understanding of the need to account for communication needs when planning deployment~\cite{AR2021Muralidharan}. We next discuss the place of UVs in 6G networks.

\begin{figure*}
    \centering
    \includegraphics[width=7in, trim={0.0in, 0.5in, 0.1in 0.4in}, clip]{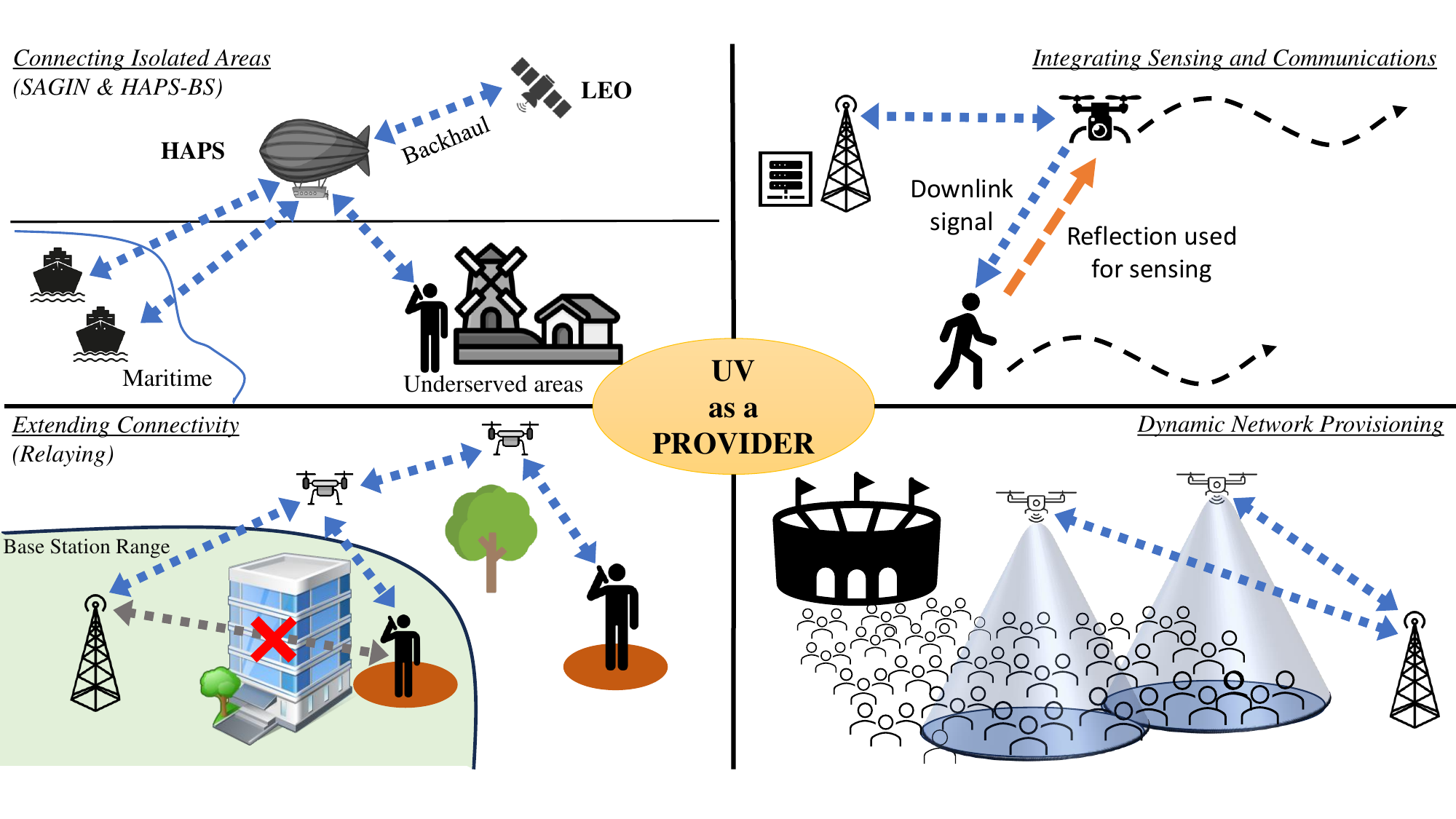}
    \vspace{-0.05in}
    \caption{Sample envisioned applications of UVs as providers in next-generation communication systems. (Top left) As part of a Space-Air-Ground Integrated Network (SAGIN), a HAPS provides connectivity to isolated and underserved areas. (Top right) A UV uses ISAC
    to simultaneously extend connectivity to and further sense end users. (Bottom left) UVs aid a base station by acting as relays. Their elevated position allows them to serve users in dead zones, and by working together, they can extend the area covered by the terrestrial base station (TBS). (Bottom right) A team of UVs provides dynamic network provisioning for a crowd during a large event.}
    \label{fig:providers_overview}
    \vspace{-0.15in}
\end{figure*}

\vspace{-0.1in}
\subsection{Unmanned Vehicles in the 6G Ecosystem}
The use of UVs can both enable and be enabled by 6G networks. \textbf{Throughout this paper, we refer to the UV as a \textit{provider} when its purpose is to aid in the realization of the promises of 6G, and we call it a \textit{consumer} when it depends on the network's capabilities to complete a mission.} 
This duality serves as an organizing principle in this paper.

\subsubsection{UVs as Providers}
Unmanned mobile infrastructure in communication systems dates back to the 1940s \cite{milivcevic2021early}, primarily for military use. Next-generation wireless networking requires adaptable mobile infrastructure \cite{dogra2020survey}, and using UVs as providers addresses many important challenges. Their mobility allows real-time network capacity provisioning to handle hotspots due to acute high-traffic events, to provide temporary solutions when permanent terrestrial infrastructure is offline, or to offer dynamic coverage to mobile users. As part of a SAGIN, HAPSs can help close the digital divide in rural areas or provide connectivity to maritime nodes, scenarios where build-out of terrestrial infrastructure is impractical \cite{CST2021Karabulut}. Furthermore, UAVs are well suited to close coverage holes (dead zones), as their elevated nature frequently creates strong line-of-sight (LOS) communication channels with ground users. Fig.~\ref{fig:providers_overview} shows samples of these provider roles.

Commercial enterprises have begun making this vision a reality. For example, in 2022, AT\&T successfully demonstrated the use of a UAV as a flying 5G base station (BS) \cite{ATT-COW}, while Aalto's Zephyr HAPS promises direct-to-device 5G connectivity with $5-10\,$ms of latency as soon as 2024 \cite{Zephyr}. As recognition of the advantages offered by UVs grows, their use will become more commonplace.

For UVs as providers, we highlight two general scenarios. First, as UV base stations (UV-BSs), they must optimize their position and allocation of limited communication resources among multiple users, while mitigating interference with users connecting via other access points. This includes using HAPS and UAVs to provision additional network capacity, to take the place of disabled terrestrial base stations (TBSs), or to extend coverage to underserved areas. Second, UVs may be deployed for cooperative communication using multihop routing or distributed beamforming. This captures problems related to extending the network or covering dead zones.

\begin{figure*}
    \centering
    \includegraphics[width=7in, trim={0.1in, 0.4in, 0.1in 0.4in}, clip]{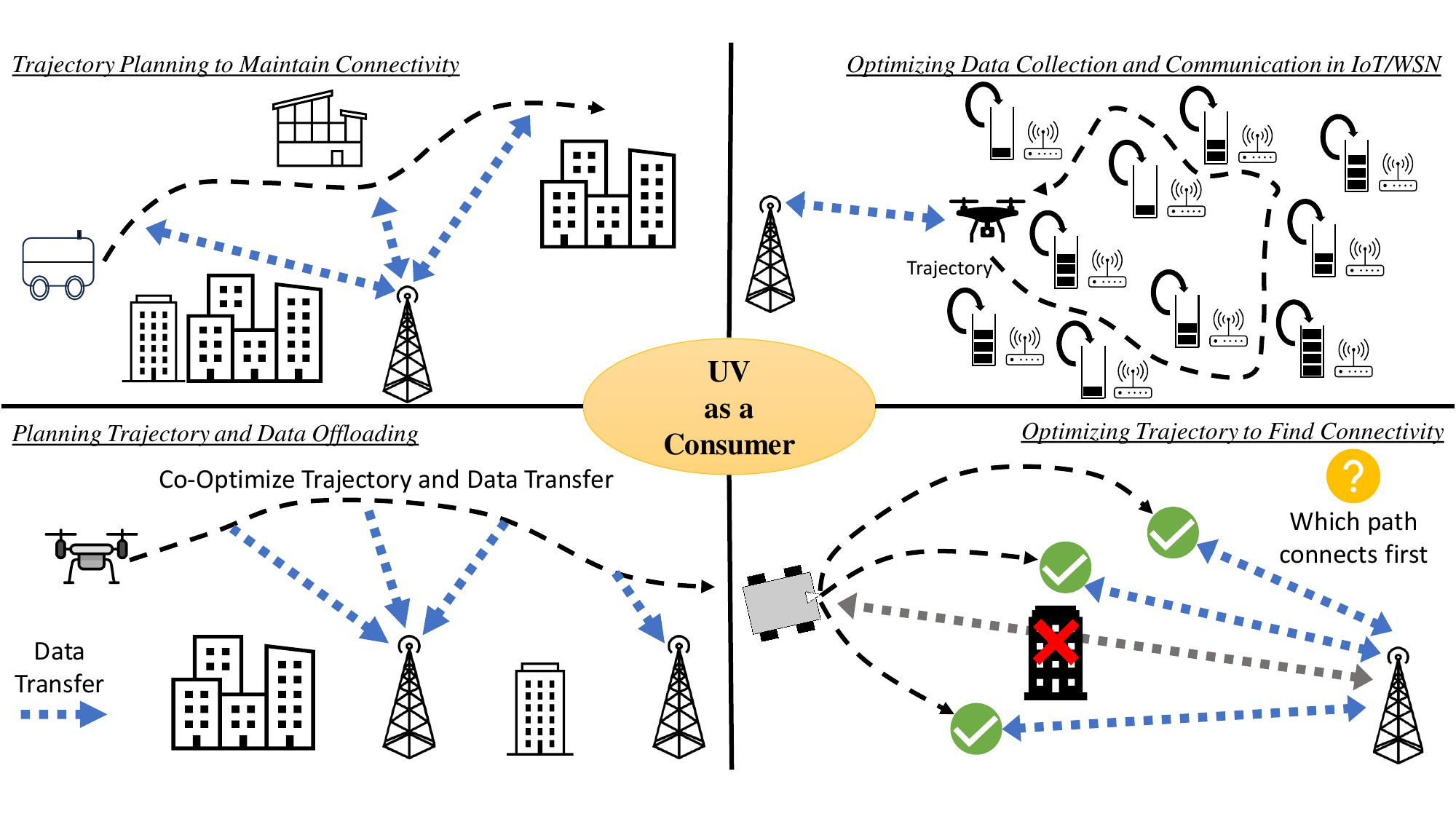}
    \vspace{-0.15in}
    \caption{Sample common tasks for UV operation in 6G networks. (Top left) A UV navigates to a destination while maintaining connectivity. (Top right)  A UV collects data from a dynamic WSN. (Bottom left) A UV must jointly plan its trajectory and transmission power/communication rate while offloading data to a remote station. (Bottom right) A UV, initially disconnected from the network, seeks to reconnect as quickly as possible.}
    \label{fig:consumer_overview}
    \vspace{-0.15in}
\end{figure*}

\subsubsection{UVs as Consumers}
On the other hand, UVs as consumers of network services will be natively supported in 6G, paving the way for their use in many exciting applications. 3GPP Release 17, published in 2022, details how 5G networks can enable UV applications \cite{3gppRel17-UAV}, with additional studies in 2024 as part of Release 18 \cite{3gpp18-UAV}. Furthermore, the Federal Aviation Administration's (FAA) Integration Pilot Program, developed to catalyze the incorporation of UAVs into U.S. airspace, published its findings in 2020 \cite{FAA-IPP}, and has since been replaced by the BEYOND program, which continues to collect flight data and develop standards towards beyond-visual-line-of-sight operation \cite{FAA-BEYOND}.

Several enterprises have already begun deploying both ground and aerial UVs for a broad range of applications. Well-known commercial examples include Alphabet's Wing \cite{Wing} and Amazon's delivery drone program \cite{AmazonDrone}, which employ cargo UAVs for last-mile delivery, while Nuro \cite{Nuro} provides a similar service with autonomous ground vehicles. Amazon's Proteus and Sequoia platforms have demonstrated the applicability of UVs in warehouse management \cite{Proteus2022Amazon}, and future industrial applications include precision agriculture and facility surveillance \cite{Access2019Shakhatreh}. In the automotive industry, Tesla \cite{Tesla}, Waymo \cite{Waymo}, and Uber \cite{Uber} lead the way towards self-driving cars, while autonomous electric vertical take-off and landing (eVTOLs) offer a vision of alternative transportation that is quickly becoming reality~\cite{namuduriSpectrum2023}. In all of these cases, communication facilitates mission completion by enabling command and control, information sharing, and access to edge computing services.

For optimal performance, UVs must jointly optimize their trajectories and communication as they complete their missions. More specifically, they must maintain connectivity for command and control, or, if connectivity is lost, they must reconnect as soon as possible. They may additionally need to schedule wireless offloading of data collected from, e.g., onboard cameras,  Internet-of-Things (IoT) devices, or wireless sensor networks, while accounting for the quality of the underlying communication channel. Data collection itself may involve deciding on the best way to visit a number of sites while taking into account the timeliness of data transfer back to a remote base station. Examples of these problems are illustrated in Fig.~\ref{fig:consumer_overview}.


\subsection{Our contributions}

Using the mobility of unmanned vehicles to enable new forms of connectivity dates back to work as early as 2008, when the term \textit{communication-aware motion planning} was coined to refer to path planning that considers its impact on communication parameters \cite{ICRA2008Mostofi}. Early work showed how a UV can actively plan its path to enable new forms of connectivity \cite{ICRA10_Mostofi_Malmirchegini_Ghaffarkhah}. This motivated subsequent work showcasing the potential of unmanned vehicles to enable new forms of connectivity \cite{AR2021Muralidharan}. While these explorations were for a general system, the possibilities and challenges offered by the 6G vision created great new interest in this area since 2019.   

Recent years have produced a number of valuable surveys and tutorials on UVs in communication systems, many in the context of 5G \cite{IoTJ2019Li, PIEEE2019Zeng, CST2019Mozaffari, VTM2020Du, SAC2021Wu, CM2021Mozaffari, CST2022Geraci, CS2023Luo}. Some of these focus on particular topics, such as machine learning \cite{VTM2020Du}, IoT networks \cite{IoTJ2019Li}, or path planning \cite{CS2023Luo}, while others provide a high-level perspective of the potential challenges in this space. 

In this context, this paper takes a unique perspective by focusing on the optimization problems that arise when planning the operation of UVs in 6G communication networks. Its primary contribution lies in the unification of the problem space, providing a structured framework to understand the use cases, underlying problem formulations, and the needed mathematical tools. Overall, the paper aims to furnish the readers with a systematic understanding of the potentials and challenges offered by the integration of UVs into the 6G communication systems.

The rest of the paper is organized as follows. In Section~\ref{sec:modeling}, we introduce the two key types of optimization variables — those related to communication resource allocation and UV mobility — and discuss key modeling considerations for each in the context of 6G. Section~\ref{sec:approaches} discusses the aspects of these systems that make optimization mathematically challenging and provides key tools for navigating this space. With this groundwork laid, Sections~\ref{sec:formulations} and \ref{sec:rl_formulations} then take a unifying perspective of the problem formulation space, methodically categorizing the large array of potential scenarios into a set of fundamental problems, while bringing key insights into relevant sub-problems. Finally, Section~\ref{sec:challenges_and_opp} gives other important considerations for the optimization of UV operation, and concluding remarks appear in Section~\ref{sec:conclusions}. \textbf{We note that Sections~\ref{sec:modeling} and \ref{sec:approaches} are written so that readers familiar with their content may choose to move directly to Section~\ref{sec:formulations}.}

%



\section{Key Concepts and Modeling}\label{sec:modeling}
The previous section highlights the many roles UVs will play in next-generation communication systems, but the realization of UVs' full potential depends on intelligent, optimized deployment of these assets in the network. In Section~\ref{sec:formulations}, we provide a unifying and structured treatment of the corresponding problem space. This section then identifies core components of these systems and presents relevant models needed for tackling the problem space. Our discussion includes established models that are still applicable for 6G, 6G-specific variations for these models, and open modeling questions. Throughout, we identify common types of constraints and performance metrics, and we end the section with a high-level problem formulation.

\vspace{-0.2in}
\subsection{Common Core Components}
To identify common system components, consider $N_b$, a bound on the number of bits that can successfully be sent between a transmitter and a receiver over a given time period, $\Delta t$. From Shannon's channel capacity for additive white Gaussian noise (AWGN) \cite{Shannon1948AMT}, this can be expressed as 
\begin{equation}\label{eq:shannon_bound}
    N_b =  \Delta t\, B\, \text{log}_{2}\big(1+\text{SINR}\big),\;\text{SINR} = {P |H|^2}/{(\sigma^2 + I)},
\end{equation}
where $B$ is the channel bandwidth, SINR is the signal-to-interference-plus-noise ratio, $P$ is the transmit power, $|H|$ is the channel gain,  $\sigma^2$ is the noise power at the receiver, and $I$ is interference power that arises from other communication links using the same physical resource block (PRB).

This bound indicates the two avenues through which UV operation impacts communication. First, $N_b$ depends on the allocation of wireless resources: the time spent communicating over the channel ($\Delta t$), the transmit power, and the bandwidth of the channel ($B$). In general, these resources are allocated among multiple users simultaneously, and the interference, $I$, may depend on allocation to other links in the network. Second, $N_b$ is a function of the communication channel, $H$, which, while generally considered exogenous, is controllable to a degree through the mobility of the UV.

Based on the discussion above, we identify three general components of these systems which merit attention. First, the communication channel affects the quality of communication over a link. Second, the allocation of communication resources across multiple users, together with the channel, determines the communication performance. Third, mobility gives the UV some control over the channel while also allowing for the completion of other mission-specific objectives. We next discuss each of these high-level components, with an emphasis on considerations for 6G and beyond.

\vspace{-0.1in}
\subsection{Communication Channel Models}\label{sec:modeling.channel}
The channel plays a central role in the performance of the system, and thus appropriate modeling is very important. However, channels display complex spatio-temporal variation, making them challenging to model and predict. We next discuss key established modeling concepts and then move to considerations for 6G and beyond. Given the large size of the frequency bands considered in 6G, the transmitted signal undergoes frequency selective fading. In the following we discuss channel models that refer to a narrow frequency band, with the understanding that the model parameters will change between frequency bands.

\subsubsection{Statistical Channel Models}
The spatial dynamics of channels are generally modeled as the superposition of three components. \textit{Path loss} describes the large-scale attenuation of the signal power as it propagates through space. \textit{Shadowing} captures the impact of large-scale blockages which create Non-LOS (NLOS) communication links. Finally, \textit{multi-path} or \textit{small-scale fading} models the small-scale variation due to multiple copies of the signal arriving at a receiver via different paths. When the multipath delay spread is relatively small, a general model has the form
\begin{equation}
    |H|^2 = H_{\text{PL},0}d^{-\alpha_\text{PL}}\; \alpha_{\text{SH}}^2 \; \gamma_{\text{MP}},
\end{equation}
where $H_{\text{pl},0}$ is the \textit{path loss intercept} (\textit{i.e.}, the path loss at a reference distance of $1\,$m), $\alpha_\text{PL}$ is the \textit{path loss exponent}, $d$ is the distance between the transmitter and the receiver, $\alpha_\text{SH}^2$ is the shadowing power, and $\gamma_{mp}$ is the multipath effect. While the path loss intercept and exponent are modeled as scalar values,  shadowing, $\alpha_{\text{SH}}^2$, is generally modeled as a spatially correlated log-normal random variable, and multipath effects can be modeled with Nakagami, Rayleigh, Rician, or log-normal distributions \cite{WirelessComms2005Goldsmith, TWC2011Malmirchegini}. The values of the path loss intercept and path loss exponent parameters, as well as the parameters for the shadowing and multipath distributions, depend on the carrier frequency and environmental factors. For wideband channels where delay spread needs to be considered, as will frequently be the case in 6G, the same modeling principles can be extended over the delay dimension, resulting in the overall power-delay profile of the channel. 

\subsubsection{Temporal Channel Variation}
Importantly, communication channels may vary across time as well as space, due to changes in the environment. To account for this, a block-fading model assumes the channel is temporally stationary over a coherence time of length $\Delta_C$ (see, \textit{e.g.}, \cite{JWCN2019Kim}). Alternatively, the shadowing power over both space and time may be modeled with a Gaussian process with an autocorrelation function
\begin{equation}\label{eq:spatio_temproal_AR}
    R\big(\alpha_\text{SH}(q,t), \alpha_\text{SH}(q',t')\big)=\alpha_\text{SH}^2 e^{-\frac{||q-q'||_2}{d_c}-\frac{|t-t'|}{t_c}},
\end{equation}
where $d_c$ and $t_c$ give the rate of decorrelation in space and time, respectively \cite{TVT2010Oestges}.

\subsubsection{Channels for UAVs and HAPSs}
For UVs deployed at high altitudes, the shadowing and multipath components may be omitted to model scenarios of unimpaired propagation between nodes, both for air-to-air links between UVs and for air-to-ground links when the height of a HAPS or UAV results in a dominant LOS link \cite{QualcomLOS}. 

For air-to-ground channels, shadowing is often accounted for by using different path loss intercepts and exponents depending on whether a dominant LOS path exists between the transmitter and receiver:
\begin{equation}\label{eq:LOS_NLOS}
   |H|^2 =\begin{cases}
       H_{\text{PL, L},0}||q_i - q_j||_2^{-\alpha_\text{PL, L}}&\hspace{-0.1in}\text{w/ prob.}\, p_{\text{LOS}}\\
        H_{\text{PL, N},0}||q_i - q_j||_2^{-\alpha_\text{PL, N}}&\hspace{-0.1in}\text{w/ prob.}\, (1 - p_{\text{LOS}})
   \end{cases},
\end{equation}
where $H_{\text{PL, L},0},\,\alpha_\text{PL, L}$ and $H_{\text{PL, N},0},\,\alpha_\text{PL, N}$ are the path loss intercept and exponent for the LOS and NLOS conditions, respectively, and $p_\text{LOS}$ is the probability of the existence of the LOS path. 

Three common models for calculating $p_\text{LOS}$ appear in the literature. A method used by the ITU relies on three parameters to statistically characterize the probability of a LOS channel existing between a UAV and a ground node \cite{ITUP1410-5}. The parameters of this model are $\beta$, the number of buildings per square kilometer, $\alpha$, the fraction of area covered by buildings, and $\kappa$, the parameter for a Rayleigh distribution describing the building heights. The formula is given by 
\begin{equation*}
p_{\text{LOS}}=\prod_{n=0}^{\max\left(0,o-1\right)}\left[1-\exp\left(-\frac{\left(h_{\text{gr}}+\frac{\Delta h\left(n+0.5\right)}{o}\right)^{2}}{2\kappa^{2}}\right)\right],
\end{equation*}
where $h_{\text{gr}}$ is the height of the ground node, $o=\lfloor \sqrt{\alpha \beta} \rfloor$ with $\lfloor\cdot \rfloor$ the floor operator, and $\Delta h$ is the difference in height between the UAV and the ground node. 

Alternatively, 3GPP has produced recommendations TR36.777~\cite{3GPPTR36777} and TR36.873 \cite{3GPPTR36873}, which calculate $p_\text{LOS}$ as:
\begin{equation*}
p_{\text{LOS}}=
\begin{cases}
1,  \hspace{0.1in} \text{if}\,d_{\text{horiz}}\leq d_{1}\, \text{or}\, h_u \geq 100\\ 
\frac{d_{1}}{d_{\text{horiz}}} +
\left(1-\frac{d_{1}}{d_{\text{horiz}}}\right)\text{exp}\left(\frac{-d_{\text{horiz}}}{d_{2}}\right), \hspace{0.1in} \text{if}\; d_{\text{horiz}} > d_{1}
\end{cases},
\end{equation*}
where $h_u$ is the height of the UAV, $d_{\text{horiz}}$ is the horizontal distance between the UAV and the ground node, $d_1=\text{max}\{460\text{log}_{10}h_u -700, 18\}$, and $d_2=4300\text{log}_{10}(h_u)-3800$. Importantly, this model holds for $22.5\,\text{m}<h_u\leq 300\,\text{m}$. 

Finally, work in \cite{WCL2014Al-Hourani, GLOBECOM2014Al-Hourani}, based on the ITU model, finds that in many scenarios, the probability of a LOS channel is modeled well as
\begin{equation}\label{eq:plos_alhourani}
p_\text{LOS}=\frac {1}{1+b_1 \exp (-b_2 (\theta-a))},
\end{equation} 
where $b_1$ and $b_2$ are environment parameters, and $\theta$ is the UAV's angle of elevation with respect to the ground node. 

Additional details on air-to-ground communication channels may be found in \cite{Access2019Yan, CST2019Khawaja, khuwaja2018COMST}.

\subsubsection{Communication Channels at mmWave and Beyond} While the modeling discussed thus far suffices for a wide range of frequencies, modeling 6G frequencies (mmWave and THz) requires additional considerations. Modeling these channels continues to be an area of active research, with details found in \cite{CST2018Hemadeh, CST2022Han}.

\textit{Acute Signal Attenuation:} 
mmWave and THz communication signals attenuate much more quickly as they travel through space, due to the physics of electromagnetic propagation as well as increased impacts of atmospheric absorption, necessitating the use of beamforming for energy-efficient communication. Furthermore, signals at these frequencies do not propagate well through most solid material due to acute penetration loss. Consequently, the existence of a LOS path between the transmitter and receiver becomes more important, though technologies, such as RIS, which can passively steer incident electromagnetic waves around blockages, may mitigate this challenge.

\textit{Sparse Multipath:} In part due to acute signal attenuation, 6G frequencies tend to be multipath sparse, so the channels tend to be dominated by path loss and shadowing.

\subsubsection{Channel Prediction}
For fixed transmitters and receivers, there are many established techniques to estimate the channel gain, but for trajectory planning, it is often important to know the channel gain at points not yet visited by the UV. The UV may employ statistical modeling for channel prediction across the workspace. For example, given a few prior channel measurements, \cite{TWC2011Malmirchegini} proposes a Gaussian-process based channel predictor which can predict the probability distribution of the channel gain at any unvisited location. The framework includes both path loss and shadowing issues, which are relevant to 6G systems. \cite{TSP2022Evmorfos} extends the Gaussian process model across space and time for predicting spatio-temporal variations. The work of \cite{Kalogerias-IEEETSP-2018Spatially} proposes a distributed stochastic optimization method to predict the Signal-to-Noise Ratio (SNR) at future time steps for mobile relay motion control, assuming that the log-magnitude of the channel evolves as a Gaussian process that is correlated across time and space along the lines of \eqref{eq:spatio_temproal_AR}. The work by \cite{Diamantaras-MLSP-2019Optimal} proposes a correlated bandits approach to predict the spatio-temporally correlated channels. In \cite{karanam2023SJ}, the authors demonstrate how the mobility of a UV can be used to facilitate full ray makeup channel prediction, enabling the prediction of the magnitude, delay and angle of arrival of all the incoming paths at unvisited locations.

More recently, Machine Learning (ML) has gained importance for channel prediction, particularly in the context of 6G~\cite{becvar2023CommsMag}. For example, \cite{Asilomar2022Torun} uses a convolutional deep neural network to predict channel values across the workspace with only a few prior measurements and without knowledge of the transmitter location.

\subsubsection{Ray Tracing} Ray tracing models communication channels by tracing the path of electromagnetic radiation from the transmitter to the receiver, explicitly accounting for phenomena such as reflection, diffraction, refraction, and scattering by splitting, attenuating, and redirecting the rays. This approach requires detailed information about the environment, and for all but the simplest cases needs computationally-intensive simulation. As a result, these methods have generally been used in offline settings, with applications, such as Remcom's Wireless InSite \cite{remcom} and Nvidia's GPU-accelerated Siona ray tracer \cite{Hoydis2023GLOBECOM_siona}, capable of producing high-fidelity simulated channels. However, recent developments in mobile hardware, such as Qualcomm's Snapdragon 8 Gen 2 \cite{Snapdragon8Qualcomm}, have raised the prospect of real-time ray tracing on small, energy-limited devices. This, in conjunction with real-time data from UV-mounted vision systems, may eventually permit the use of digital twins when planning 6G network operations, an idea which has generated much excitement recently~\cite{alkhateeb2023CommsMag}.

\subsubsection{Realistic Channel Environments}
Regardless of which channel model is used, good modeling and validation require grounding in real-world measurement campaigns. Standards and recommendations written by bodies such as the ITU and 3GPP are good resources for finding realistic parameters for statistical channel modeling \cite{ITUP1410-5, 3GPPTR36777, 3GPPTR36873}, as are academic works which take on the challenge of measurement campaigns, several of which are summarized in \cite{CST2019Khawaja}.

Alternatively, realistic channel data may be synthesized. For example, \cite{mostofi_lab_channel_simulator} generates statistically realistic channel spatial variations using a Gaussian process model for shadowing (in dB domain), with an exponential correlation function, as discussed earlier, while also accommodating multipath effects if needed, based on the work in \cite{JRobotics11}, and \cite{NYUSIM} provides sophisticated channel simulation for 5G and 6G mmWave and sub-THz channels in environments studied by 3GPP. Similarly, the work of \cite{kalogerias2017spatially} (as outlined in Theorem 2) shows how to synthesize channels that have the spatio-temporal characteristics of \eqref{eq:spatio_temproal_AR} based on a stable order-1 vector autoregression, with a corresponding software package found in \cite{evmorfos2021ReinforcementCode}. As noted earlier, when sufficient environmental information is available, ray tracing software may also be used.

\subsubsection{Model-Free Channels} While we have focused on analytical or simulated models, data-driven, model-free approaches offer an interesting avenue to capture channel characteristics, particularly in the context of reinforcement learning (RL) \cite{ICASSP2022Evmorfos}. A key challenge here lies in undertaking measurement campaigns large enough to capture channel variations across both space and time.

\vspace{-0.1in}
\subsection{Communication Essentials}
\label{subsec:metrics}
For a single communication link, the allocation of limited power and bandwidth over a finite time, together with the channel, determine the performance of the communication system, as indicated by the bound in \eqref{eq:shannon_bound}. When multiple links are considered, the picture becomes more complex due to the possibility of interference and the need to aggregate the performance over each link. We next discuss interference management via multiple access schemes, focusing on emerging  multi-access techniques for 6G, before reviewing important metrics of the communication system's performance.

\subsubsection{Multiple Access Schemes}
Multi-access schemes define resource allocation rules which mitigate interference between users connecting to the same BS. Traditionally, orthogonal multiple access (OMA) schemes have been used to ensure that no two users send or receive data on the same PRB by splitting either time (time-division multiple access (TDMA)), or bandwidth (orthogonal frequency-division multiple access (OFDMA)) into equal, non-overlapping partitions of a predetermined size and assigning a single user to each.

To achieve the objectives of 6G, physical resources must be used more efficiently, leading to increased interest in non-orthogonal multiple access schemes, which allow multiple channels to occupy the same PRB. For example, building on classical multiple-input multiple-output (MIMO) techniques,  massive (mMIMO) has been a key enabler of 5G, allowing for space-division multiple access (SDMA) by creating strong antenna gains in multiple directions simultaneously \cite{FNWF2022Sun}. In 6G, \textit{ultra-massive} MIMO \cite{samsung_6g} or \textit{extreme} MIMO \cite{IMT-2030} will continue to develop this approach. Alternatively, PRBs can be reused by employing intelligent signal processing and interference management with techniques such as power-domain Non-Orthogonal Multiple Access (PD-NOMA) \cite{JNCA2022Ghafoor} and RSMA \cite{10038476}. 

In the past few years, orthogonal time frequency space (OTFS) modulation~\cite{hadani2017WCNC_OTFS} has also emerged as a strong candidate for 6G waveforms due to its ability to overcome challenges caused by strong Doppler shifts. These pronounced Doppler shifts can destroy the orthogonality of OFDMA, the preferred modulation method of 5G, but OTFS-based multi-access schemes~\cite{khammammetti2019WCL} offer a way to achieve the level of mobility envisioned for 6G~\cite{IMT-2030}, especially in mmWave and higher frequencies~\cite{hadani2017IMS_OTFS}. In OTFS, the channel is considered in the Delay-Doppler domain, where  a time varying channel is sparsely represented and appears linear and time-invariant.
 
\subsubsection{Communication Metrics}
\label{subsec:utility_metr}
In IMT-2030, the ITU has set out a variety of metrics, both new and long-standing, to measure the performance of 6G systems, as we review next.

Fundamentally, the SINR of individual communication links determines the performance of the network. The SINR in turn determines the maximum achievable spectral efficiency, $r^{\text{max}} = \text{log}_{2}(1+\text{SINR}),\,$ and by extension the maximum achievable communication rate. The performance may also be measured in terms of energy efficiency, \textit{i.e.}, bps/W. 

On the other hand, by modeling data generation at an end user as an $M/G/1$ queue, the latency across a single link is
\begin{equation*}
    l = \frac{\lambda}{2 (B\,r^\text{max}\nu)((B\,r^\text{max}/\nu) - \lambda)} + \frac{\nu}{Br^\text{max}}, 
\end{equation*}
where $\nu$ is the packet size \cite{TWC2019Challita} and $\lambda$ is the data generation rate for user $u$ in bps. 

Another key communication metric is the bit-error rate (BER), which is the ratio of the number of bit errors to the total number of transmitted bits. The BER is calculated in various ways depending on factors such as modulation scheme, coding scheme, and channel characteristics \cite{WirelessComms2005Goldsmith}. Assuming M-ary quadrature amplitude modulation (M-QAM) communication constellation with spectral efficiency $r$, the reliability in terms of the BER is well approximated by $p_\text{b}\hspace{-3pt} \approx \hspace{-3pt} 0.2\text{exp} \left(-1.5 \text{SNR}/\left(2^r\hspace{-2pt} -1\right)\right)$ for certain SNR regimes~\cite{WirelessComms2005Goldsmith}.

Other important metrics include coverage, outage probability, and the number of handoffs experienced by a user, i.e., the number of times the user's BS association changes, while secrecy rate provides a way of evaluating the security and privacy of the system \cite{PLS2011Bloch}.

\subsection{UV Mobility: Kinematics and Energy Models}
So far, we have discussed key communication concepts and models. We now address another important facet, the positioning and movement of UVs. Mobility supports UVs both as providers and consumers in the network by giving some degree of control over the communication channel, while also allowing for the completion of mission-specific tasks such as data harvesting or automated delivery. We next briefly present key considerations for modeling the mobility of UVs including kinematic and energy consumption models.

\subsubsection{Dynamics and Kinematic Constraints}
A basic model treats the UV as a single point, $q$, constrained to lie in some region $\mathcal{Q}\in\mathbb{R}^3$, which may account for obstacles or no-fly zones. In the multi-UV case, a minimum distance constraint between UVs may enforce collision avoidance.

A simple second-order dynamical system provides a basic model of mobility, expressed as 
\begin{equation}
    \dot{q} = v,\;\;\dot{v} = u,
\end{equation}
where the dimensions of velocity, $v$, and control input (acceleration), $u$, match the dimension of $q$. When planning a 2D trajectory, either as a ground vehicle or when flying at a fixed altitude, the UV's velocity, $v$, is also constrained to be in some range, \textit{i.e.}, $v_\text{MIN} \leq \|v\| \leq v_\text{MAX}$, with the minimum and maximum values dependent on the specifics of the UV. For ground vehicles and rotary-wing UAVs, $v_\text{MIN} = 0$, while fixed-wing UAVs will have some non-zero minimum flight speed. For 3D trajectories, the velocity constraint may be further decomposed into two separate vertical and horizontal maximum velocities, as for many UAVs, the mechanics of moving vertically differ significantly from those of moving horizontally \cite{TWC2020You, dji_phantom}. Acceleration should be similarly bounded to capture realistic kinematic constraints.

This basic model has the advantage of simplicity and serves when planning for rotary-wing  UAVs or, more generally, when the UV has a small turn radius compared to the scale on which the planning occurs. At other times, however, more detailed kinematics are important. For many ground vehicles, a basic car model offers a simple yet descriptive representation \cite{PlanningAlgos2006LaValle}. This differential model includes not only the position of the vehicle, $q$, but the orientation $\theta_o$ as well. The system takes as input a velocity, $v$, and steering angle, $\phi$, and the dynamics are
\begin{equation}
    \dot{q} = v[\text{cos}\theta_o,\,\text{sin}\theta_o],\,\dot{\theta_o} = \phi_o,
\end{equation}
with $\dot{q} = [\dot{q}_{x}, \dot{q}_{y}]$ denoting the velocity in the x- and y-coordinates. Generally, the steering angle is constrained, i.e., $|\phi_o|\leq \phi_{o,\text{MAX}}$, and the minimum turning radius is given by $L/\text{tan}\phi_{o,\text{MAX}}$, with $L$ as the distance between the two axles. If $v$ is allowed to be negative, the vehicle may go in reverse, but when the velocity is constrained so that $v\in\{0,1\}$, we have the well-studied Dubins model \cite{AJM1957Dubins}.

The kinematics of fixed-wing aircraft are quite complex, but a simple extension of the model above gives an elementary yet useful set of dynamics. In particular, the dynamics at a fixed flying altitude and constant velocity can be modeled directly as a Dubins model, while any changes in altitude can be captured by the independent linear system $\dot{q}_z = v_z$, for a vertical velocity $v_z$ \cite{PlanningAlgos2006LaValle}.

The above models give a more accurate sense of a UVs movement capabilities than the simple point model first presented. However, some scenarios, such as when studying air-frame occlusion or beam-misalignment, require rigorous understanding of UV kinematics \cite{CL2023Bonilla, CST2019Khawaja}. In these cases, appropriate models must be found using established techniques from robotics.

\subsubsection{UV Energy Models}
UVs generally have access to limited energy, so understanding the energy requirements for a proposed trajectory is critical to successful operation. As with kinematics, exact expressions for energy consumption are tied to the specifics of the UV under consideration, but some general models capture important features for different types of UVs. For example, \cite{TR2006Yongguo} develops an empirical model for a pioneer ground robot given by $P(v) = \kappa_1 v + \kappa_2$, where $\kappa_1$ and $\kappa_2$ are determined by the load and mechanics of the robot.

Alternatively, \cite{TWC2019Zeng} gives the instantaneous power consumption for a rotary-wing UAV as a function of velocity with the following formula:
\begin{align}\label{eq:rotary_UAV_energy_model}
P(v)=&\underbrace {P_{0} \left ({1 + \frac {3v^{2}}{U_{\mathrm {tip}}^{2}} }\right)}_{\text {blade profile}}+ \underbrace {P_{i} \left ({\sqrt {1 + \frac {v^{4}}{4 v_{0}^{4}}}-\frac {v^{2}}{2v_{0}^{2}}}\right)^{1/2}}_{\text {induced}} \\&+\,\underbrace {\frac {1}{2} d_{0} \rho s_r A v^{3}}_{\text {parasite}},\notag
\end{align} 
where, $v$ is the velocity, $P_0$ and $P_i$ are constants, $U_\text{tip}$ is the speed of the tip of the rotor blade, $v_0$ is the mean rotor induced velocity in hover, $d_0$ is the fuselage drag ratio, $s_r$ is the rotor solidity, $\rho$ is the air density, and $A$ is the rotor disk area. Importantly, this model is not convex in $v$, but a convex approximation fits well outside the low-velocity regime.

Finally, for a fixed-wing UV flying at a fixed altitude, \cite{TWC2017Zeng} derives an energy model which accounts for drag, list, and thrust. For the velocity vector $\mathbf{v}\in \mathbb{R}^2$ and its time derivative, $\dot{\mathbf{v}}$, the expression is given as:
\begin{align}\label{eq:fw_UAV_energy_model}
P(\mathbf{v}, \dot{\mathbf{v}})= c_1 \|\mathbf{v}\|^3+\frac{c_2}{\|\mathbf{v}\|}(1+\frac{\|\dot{\mathbf{v}}\|^2 - \frac{\dot{\mathbf{v}}^T \mathbf{v}}{\|\mathbf{v}\|^2}}{g^2}) + \frac{m\|\dot{\mathbf{v}}\|}{4},
\end{align} 
where $m$ is the mass of the UV, $g$ is acceleration due to gravity, and $c1$ and $c2$ are constants determined by other parameters of the UV air frame. The $1/\|\mathbf{v}\|$ and $\|\mathbf{v}\|^3$ terms capture high power consumption at both very low and very high velocities.

Total onboard energy clearly imposes a hard constraint on operation. This may motivate connecting UVs to power sources via tethers in some cases, which will then impose some additional kinematic constraints \cite{khemiri2023OJCOMS, 9515754, 9921200}. In most envisioned cases, however, the UV can periodically return to a location to charge or swap batteries~\cite{cicek2024TITS, guo2020ICC}, or its energy budget and charging schedule can be optimized as part of the whole operation, as we shall discuss in the rest of the paper.

\subsection{A High-level Problem Formulation}
To conclude, we present a high-level problem formulation consisting of an objective — which may account for the communication performance/cost, motion costs, and other mission objectives — and constraints, which can be motivated by communication, motion, or mission-specific considerations. The decision variables can be broadly categorized as either communication- or trajectory-related, with the possible addition of mission-related variables representing concepts such as sensing. Let $\mathcal{R}$, $\mathcal{T}$, and $\mathcal{M}$ represent collections of variables that determine the allocation of communication resources, the UV's positioning, and other mission-specific optimization variables, respectively. Then a high-level formulation is shown in (\ref{eq:generic_opt}),

\begin{maxi}|s|[2]                   
    {\substack{\mathcal{R}, \mathcal{T},\\ \mathcal{M}}}                               
    { f_\text{GLB}\big( \underbrace{f_{\text{com}}(\mathcal{R}, \mathcal{M})}_{\substack{\text{Communication}}},\,\underbrace{f_{\mathcal{K}}(\mathcal{T}, \mathcal{M} )}_{\text{Motion}},\, \underbrace{f_m(\mathcal{R}, \mathcal{T}, \mathcal{M})}_{\text{Mission}} \big) }   
    {\label{eq:generic_opt}}             
    {}                                
    \addConstraint{ \hspace{-0.3in}   \begin{rcases}
    &g_{\text{com}}(\mathcal{R}, \mathcal{M}) \leq 0\\
    &h_{\text{com}}(\mathcal{R}, \mathcal{M}) = 0
    \end{rcases}\,\text{Communication constraints}
    }  
    \addConstraint{ \hspace{-0.3in}   \begin{rcases}
    &g_{\mathcal{K}}(\mathcal{T}, \mathcal{M}) \leq 0\\
    &h_{\mathcal{K}}(\mathcal{T}, \mathcal{M}) = 0
    \end{rcases}\,\text{Motion constraints}
    } 
    \addConstraint{  \hspace{-0.3in}  \begin{rcases}
    &g_{m}(\mathcal{R}, \mathcal{T}, \mathcal{M}) \leq 0\\
    &h_{m}(\mathcal{R}, \mathcal{T}, \mathcal{M}) = 0
    \end{rcases}\,\text{Mission constraints}
    } .
\end{maxi}

The objective may include communication, motion, and mission terms, combined as a weighted sum or in ratios. Not all terms will be present in all formulations, and furthermore, performance in any one of these areas may also be captured as performance threshold constraints.

Communication constraints come in two types. First, there are communication resource constraints, which ensure that time, bandwidth, or spatial degrees of freedom are not over-allocated. Second, there are constraints tied to communication related metrics, including constraints on the minimal latency, SNR, or outage probability. Importantly, when a stochastic channel model is used, the constraints must be defined either in expectation or in some other probabilistic sense (\textit{e.g.}, the constraint is met with some probability).

Motion constraints can also be subdivided into a few categories. First, there are kinematic constraints related to what movement is physically possible by the UV. Second, there may be safety constraints, such as a minimal flying height for a UV or a minimum distance between any two UVs. Finally, there may be constraints related to energy consumption for motion.

Finally, the mission-specific problem may introduce additional constraints or objectives. For example, if the UV is tasked with visiting a number of sites, we may additionally have a constraint that captures this requirement, or the objective may be to do so in the minimum amount of time, subject to, \textit{e.g.}, energy and power constraints. Other examples include UV fleet size, budget, and sensing quality constraints.

\section{An Optimization Toolkit}\label{sec:approaches}
In the previous section, we highlighted key modeling considerations and developed a high-level problem formulation which captured the many different optimization problems that arise when considering the operation of UVs in next-generation communication networks. However, it is difficult to develop a similarly generic solution method due to the wide variety of specific problem instances represented by the generic formulation. Rather, this section presents a number of important tools which find broad application, and these are classified into two categories. The first category comprises analytical methods, which generally produce interpretable solutions with well-defined optimality properties. Acknowledging the importance of AI and ML in the 6G vision, the second category encompasses learning-based approaches, particularly in the context of stochastic dynamic programming with stochastic function approximation. These powerful methods are capable of solving otherwise-intractable problems at the expense of interpretability of the solution.

\subsection{Analytical Methods}

Below, we discuss facets of UV operation that add complexity to problem formulations and highlight useful analytical tools for tackling them. We draw from fields such as computer science, operations research, robotics, and control, where many optimization tools have been developed into well-understood frameworks with clear understanding of optimality guarantees and computational complexity. 

\textbf{Handling Non-convexity with Convex Approximation:}
In many instances, meaningful measurement of performance requires consideration of nonlinear objectives or constraints (see Section \ref{sec:modeling}). For example, spatially non-convex channel gains or non-convex energy models, such as that of Eq.~(\ref{eq:rotary_UAV_energy_model}), create non-convexity in trajectory design problems. 

For non-convex objective functions, \textit{successive convex approximation} (SCA) provides an iterative method that guarantees a locally optimal solution under permissive technical assumptions \cite{TSP2014Scutari}. Specifically, for an objective function $f(x)$, the method finds a sequence of intermediate solutions $\{x_k\}$ by solving a series of simpler problems, $x_{k+1} = \argmin_{x\in\mathcal{X}} \tilde{f}(x|x_{k})$, where $\mathcal{X}$ is the solution space and $\tilde{f}(x|x_{k})$ is a strongly convex local approximator of $f$, possibly parameterized by the current iterate $x_k$. Newton's method is a common example of SCA, in which $\tilde{f}(x|x_{k})$ is the second-order Taylor-series expansion around the current iterate, $x_k$. 

In some instances, coupling between optimization variables makes the problem difficult to solve. In such cases, \textit{block coordinate descent} (BCD) simplifies the problem by partitioning the variables into non-overlapping subsets (blocks) and iteratively optimizing over a single block at a time. When the problem is convex in the block variables, standard convex optimization methods can be used. Otherwise, methods like SCA can be used to update a single block. Various optimality results hold under certain assumptions (see \cite{Peng2023block} and references therein). For UV problems, treating positioning and communication optimization variables as distinct blocks is often a useful way to decompose the problem.

\textbf{Handling Combinatorics with Mixed Integer Programs:}
Modeling concepts such as user-BS association or visit sequence among a set of Wireless Sensor Network (WSN) cluster heads introduces binary variables, giving these problems a combinatorial aspect. Mixed integer programs (MIPs), a broad class of mathematical programs characterized by the presence of both continuous- and integer-valued variables, can model these scenarios. While the general MIP is NP-hard, for special cases, extensive literature offers established methods for finding high-quality solutions. 

Within the umbrella of MIPs, there are several important subclasses of problems characterized by the nature of the objective and constraint functions. In mixed-integer linear programs (MILPs), the objective and all constraint functions, $f_i$, are linear, whereas in a mixed-integer nonlinear program (MINLP)\cite{MINLP2012Lee}, the objective and/or constraints may be nonlinear. Within the class of MINLPs, we further have a mixed-integer quadratic program (MIQP) \cite{MIQP2017DelPia} if the objective is quadratic and the constraints are linear, and a mixed-integer second-order cone program (MISOCP) \cite{MISCOP2013} if the objective is linear and all constraints are second-order cone constraints.

Branch-and-cut or branch-and-bound methods are common candidates for finding solutions of MIPs. These methods iteratively tighten upper or lower bounds on the optimal solution via relaxations of the integer variables or the nonlinear constraints/objective. In the case of MILPs, the resulting relaxations are linear programs (LPs), which can be solved very quickly, so that optimal solutions may be found quite efficiently. Similarly, the convexity of the relaxations of MIQPs and MISOCPs allow reasonably sized problem instances to be solved optimally in a modest amount of time. Several commercially available software packages are capable of solving these types of problems, including IBM CPLEX \cite{IBM_CPLX}, MOESK \cite{MOSEK}, and GUROBI \cite{GUROBI}. For general MIPs, however, finding optimal solutions is quite challenging, and finding high-quality approximate solutions is often acceptable.

\textbf{Finding Visit Orders Using Vehicle Routing Problems:} The vehicle routing problem (VRP) \cite{VRP2022Toth} refers to a general set of problems, including the classical traveling salesperson problem (TSP), in which one or more vehicles must visit a number of locations in the shortest time possible. These problems may be posed as MIPs, with integer variables giving the sequence of site visits. Time window, energy, and vehicle-capacity constraints may be included as well. 

The core modeling concept of the VRP is a weighted graph, with nodes representing locations to visit and edge weights giving the cost to move between sites. Works such as \cite{TxITS2022Pasha, COR2020Chung, EMMP2016Agatz} provide an overview of the application of these types of problems for UV trajectory design. This framework can be made communication aware by incorporating communication-aware metrics into the edge weights.

\textbf{Solving Shortest Path Problems with Graph-Theoretic Algorithms:}
Many problems of interest can be cast as shortest-path problems, particularly for scenarios where the UV is a consumer of network connectivity. For example, a cargo UAV may need to find the shortest path to its destination while respecting a quality-of-service (QoS) constraint or minimizing handoffs across BS. In these cases, the space may be discretized and modeled abstractly as a graph, and $\text{A}^*$ \cite{TSSC1968Hart} or similar graph-based algorithms can produce optimal solutions.

Randomized methods from the robotics community provide interesting alternatives. The Rapidly-exploring Random Tree algorithm (RRT) finds a path between a given start and end positions by building out a tree of partial solutions constructed from random samples of the UVs configuration space. This method quickly finds feasible solutions, then iteratively improves, making it a strong candidates for real-time applications, where available time for computation is unknown and short. Furthermore, it naturally handles obstacle avoidance as well as kinematic constraints, and a variant, $\text{RRT}^*$, provides asymptotic optimality guarantees \cite{IJRR2011Karaman}.

\textbf{Sequential Decision Making with Dynamic Programming and Optimal Control:}
Both the VRP and shortest-path problems are examples of more general sequential decision-making problems, for which dynamic programming (DP) and optimal control provide highly developed frameworks in discrete- and continuous-time settings, respectively. While some DP problems can be solved via analyzable and provably-optimal methods such as value iteration, many problems necessitate a learning-based approach. We discuss these methods further in the next subsection, but before doing so, we highlight the use of analytical optimal control.

When the dynamics of the problem are described with ordinary differential equations (ODEs), \textit{e.g.}, when using a more developed kinematics model, optimal control \cite{OptCntrl2011Liberzon} offers an established framework for planning the UV's operation. Two key entities must be well-defined before posing the optimal control problem. First, an ODE in the form  $ \dot{x}(t) = f(x(t), u(t))$ describe the evolution of the trajectory, $x(t)$, over time, with different control inputs, $u(t)$, producing different trajectories. Second, a cost functional, 
\begin{equation*}
 J\left(u(t)\right) = \int_{t=t_0}^{t_f} L(x(t), u(t))\, dt\, +  K(x(t_f), t_f),    
\end{equation*}
provides a way to compare the quality of different inputs, where $L$ and $K$ are functions capturing the instantaneous and terminal costs/rewards, respectively. 

As discussed in \cite{DPandOC2017Bertsekas}, optimal control can be conceptualized as DP in continuous time. Analogous to the Bellman equation, the Hamilton-Jacobi-Bellman equation characterizes the optimal control policy, and when the cost functional has special structure, more specific results may be applied. For example, if $f(x(t), u(t))$ describes a linear time-invariant system and $L(x,u)$ can be expressed in the quadratic form $x^T Q x + u^T R u + 2x^T Nu$ where $Q$ and $R$ are symmetric positive-definite matrices, and if there are no terminal costs, i.e., $K(x,u) = 0$, we have a linear-quadratic regulator (LQR) problem, for which solutions may be found by solving an algebraic Riccati equation \cite{LST2009Hespanha}.

\subsection{Learning-based Methods}\label{sec:approaches.DRL}

\textcolor{black}{Trajectory design for UVs often requires optimizing system parameters, addressing long-term operation, and managing stochastic uncertainties. Stochasticity is inherent in this domain, particularly due to difficult-to-predict spatial and temporal channel variations brought about by shadowing and multipath effects \cite{shaikh2018robust}. Joint mobility of end-users and UVs further induces additional uncertainty. By taking a data-driven perspective and including stochastic function approximation, learning-based methods provide for the optimization of UV-assisted communication systems while ensuring robustness against uncertainty \cite{feriani2021robustness}.} 

\textcolor{black}{In this context, the reinforcement learning (RL) paradigm  \cite{sutton-MITPRESS-2018reinforcement, bertsekas-AthenaScien-2019reinforcement} considers an agent that seeks to learn a policy for choosing actions by interacting with the environment. The agent's goal is to maximize a notion of cumulative reward over time, more specifically, the expected sum of rewards that the agent collects over a horizon of interactions.}

\textcolor{black}{Mathematically, the RL problem can be formulated as a Markov Decision Process (MDP). An MDP is defined by a tuple $ (\mathcal{S}, \mathcal{A}, \mathcal{P}, R, \gamma) $, where:}
\textcolor{black}{\begin{itemize}   
\item $ \mathcal{S} $ is the set of states representing the environment's observable conditions.
\item $ \mathcal{A} $ is the set of actions that the agent can take to interact with the environment.
\item $ \mathcal{P}(\mathbf{s}_{k+1}|\mathbf{s}_k, \mathbf{a}_k) $ is the transition function, specifying the probability of transitioning to state $ \mathbf{s}_{k+1} $ given the current state $ \mathbf{s}_k $ and action $ \mathbf{a}_k $.
\item $ R(\mathbf{s}_k, \mathbf{a}_k) $ is the reward function that provides a scalar feedback signal to the agent after taking action $ \mathbf{a}_k $ in state $ \mathbf{s}_k $. An instantiation of the reward function at time step $k$ can be denoted as $r_{k}$.
\item $ \gamma \in [0,1] $ is the discount factor, representing the agent's preference for immediate rewards over future rewards.
\end{itemize}}

\textcolor{black}{The agent's goal is to learn an optimal policy $ \pi^* $ that maps states to actions, such that the expected cumulative reward, also known as the return, is maximized:
\vspace{-0.1cm}
\begin{equation}
   \pi^{*} = \argmax_{\pi} \mathbb{E}_{\pi}\left[\sum_{k=0}^{\infty} \gamma^{k} R\left(\mathbf{s}_{k}, \mathbf{a}_{k} \equiv \pi(\mathbf{s}_{k})\right)\right].
\end{equation} }
\vspace{-0.1cm}
\textcolor{black}{The state-action value function $Q^{\pi}$ is defined as the expected discounted sum of rewards starting from a state-action pair and following the policy $\pi$ thereafter, i.e., 
\vspace{-0.1cm}
\begin{equation}
Q^{\pi}(\mathbf{s},\mathbf{a}) = \mathbb{E}_{\pi} \Bigl[\sum_{k=0}^{\infty} \gamma^{k} R(\mathbf{s}_k, \pi(\mathbf{s}_k))| (\mathbf{s}_0, \mathbf{a}_0) = (\mathbf{s},\mathbf{a})\Bigr].
\end{equation}}
\vspace{-0.1cm}
\textcolor{black}{The optimal value function $Q^{*}(\mathbf{s},\mathbf{a})$ is the fixed point of the Bellman backup operator \cite{bellman-AAAS-1966dynamic}:
\vspace{-0.1cm}
\begin{equation}
Q^{*}(\mathbf{s},\mathbf{a}) = \mathbb{E}_{\mathbf{s}' \sim \mathcal{P}} \Bigl[R(\mathbf{s},\mathbf{a}) + \gamma \max_{a'} Q(\mathbf{s}', \mathbf{a}')\Bigr]   .
\label{bellman}
\end{equation}}
\vspace{-0.1cm}
\textcolor{black}{In deep RL (DRL) \cite{mnih-Nature-2015human}, the optimal state-action value function is often parameterized as a neural network with parameters $\mathbf{w}$, denoted as $Q_{\mathbf{w}}(\mathbf{s}, \mathbf{a})$, also known as the \textit{critic} or \textit{value network}. Approximating the value function involves using batches of transitions, typically stored in an Experience Replay $D_\text{ER}$, and applying the following gradient descent update rule on the critic:
\vspace{-0.1cm}
\begin{equation}
\begin{split}
    \mathbf{w} \rightarrow \mathbf{w} 
    +  \eta \mathbb{E}_{(\mathbf{s},\mathbf{a},\mathbf{s}',r) \sim D_\text{ER}}\Bigl[ \Bigl (&Q^{*}_{\mathbf{w}}(\mathbf{s},\mathbf{a}) - Q_{\mathbf{w}}(\mathbf{s},\mathbf{a})\Bigr)\\ &* \nabla_{\mathbf{w}}Q_{\mathbf{w}}(\mathbf{s},\mathbf{a})\Bigr] ,
\end{split}
\end{equation}}
\textcolor{black}{where $*$ is the standard multiplication operator and the scalar parameter $\eta$ denotes the learning rate.}

\textcolor{black}{In the context of  MDPs with discrete action spaces, achieving an accurate estimate of the value function is deemed sufficient for optimal control \cite{dulac2015deep}. This  arises from the fact that, at each state $\mathbf{s}_{t}$, the agent can straightforwardly select the action $\mathbf{a}_{t}$ that maximizes the value estimator. }

\textcolor{black}{Discrete action spaces are prevalent in UV problems, often resulting from the discretization of the UV's operational space, thus inducing a discrete action space MDP. Moreover, in various scenarios, the action space is inherently discrete. For instance, in situations where decisions like the UV-to-BS or the UV-to-end-user associations are made at every time step, the action space inherently takes on a discrete nature.} 

\textcolor{black}{However, when dealing with continuous action spaces, the maximization with respect to the action requires a different approach. In such a scenario, action selection is achieved by parametrizing the policy with a neural network with parameter vector $\phi$, $\pi_{\phi}(\mathbf{a}_{k} | \mathbf{s}_{k})$. The parameter vector $\phi$ is adjusted with stochastic gradient descent to choose the action at each state that maximizes the estimation of the critic
\cite{silver-ICML-2014deterministic}:
\vspace{-0.1cm}
\begin{equation}
     \nabla_{\phi} J(\phi) \approx \mathbb{E}_{\mathbf{s}_{k} \sim \mathcal{P}} \left[\nabla_{\mathbf{w}} Q_{\mathbf{w}}(\mathbf{s}_{k}, \pi_{\phi}(\mathbf{s_{k}}) )\nabla_{\phi} \pi_{\phi} (\mathbf{s}_{k}) \right].
\end{equation}}
\textcolor{black}{In UV-assisted wireless communications, decision variables like energy expenditure, motion space, and bandwidth allocation inherently take values from a continuous set. Discretizing the solution set is conventionally non-scalable due to the curse of dimensionality \cite{koppen2000curse}. Thus, a more effective strategy is to use continuous action space DRL approaches for addressing these complex problems.}

\textcolor{black}{Various versions of the policy gradient algorithm adapt gradient estimation for different purposes. Proximal Policy Optimization (PPO) \cite{schulman-Arxiv-2017proximal} employs a clipped surrogate objective to maintain stability during policy updates, preventing large changes when value estimates are noisy. Similarly, Trust Region Policy Optimization (TRPO) \cite{schulman2015trust} confines policy updates within a trust region of the parameter space to ensure smooth transitions in learning steps.}

\textcolor{black}{Certain methods avoid using a function approximator to parameterize the value function and concentrate solely on parametrizing the policy. These approaches replace the parametrization of the critic with Monte Carlo estimates of the value function, fitting within the Baseline Policy Gradient (BPG) framework. Here, the policy is often presented as a learnable distribution over the action space, and the baseline policy gradient is usually formulated as follows:}
\vspace{-0.1cm}
\begin{equation}
     \nabla_{\phi} J(\phi) \approx \mathbb{E}_{\mathbf{s}_{k} \sim \mathcal{P}} \left[\hat{R}_{k}\nabla_{\phi} \log \pi_{\phi} (\mathbf{s}_{k}) \right],
\end{equation}
\vspace{-0.1cm}
\textcolor{black}{where $\hat{R}_{k}$ is the Monte Carlo estimate of the expected sum of rewards corresponding to the state-action pair $(\mathbf{s}_{k}, \mathbf{a}_{k})$. Since Monte Carlo estimates usually exhibit high variance, the baseline policy gradient methods are considered unstable.}

\subsubsection{On-policy vs Off-policy}
\textcolor{black}{A fundamental distinction exists between \textit{On-policy} and \textit{Off-policy} DRL methods. On-policy methods optimize the current policy based on its generated data, discarding experiences from previous policies, which can slow learning. In contrast, Off-policy methods decouple data collection from optimization, using Experience Replay \cite{mnih-Nature-2015human} to store past experiences for more sample-efficient updates. While On-policy methods like TRPO \cite{schulman2015trust} and PPO \cite{schulman-Arxiv-2017proximal} prioritize stability, Off-policy methods (e.g., Deep Q-Learning (DQL) \cite{mnih-Nature-2015human} and Soft Actor-Critic (SAC)\cite{haarnoja-ICML-2018SAC}) often demonstrate greater sample efficiency and flexibility in learning from diverse experiences.}

\subsubsection{Model-based vs Model-free}
\textcolor{black}{ The dynamics of the MDP are encapsulated in the probability distribution over subsequent states and rewards given the current state-action pair $\mathcal{P}(\mathbf{s}_{k+1}, r_{k} | \mathbf{s}_{k}, \mathbf{a}_{k})$. The presence or absence of assumptions concerning these dynamics delineates the distinction between \textit{model-based} and \textit{model-free}  DRL approaches.}

\textcolor{black}{In model-based DRL, the agent has an internal representation of transition dynamics, either assumed beforehand or estimated through data-driven techniques \cite{moerland-FoundationsML-2023model, barto-Springer-2013intrinsic}. With this insight, the agent can simulate trajectories to determine the optimal policy. In model-free approaches, there is no internal representation of the underlying dynamics. Instead, the agent directly learns the value function and optimal policy through an iterative trial and error process. }

\textcolor{black}{Exploring the intersection of UV-assisted wireless communications and DRL reveals the distinction between model-based and model-free approaches, especially regarding the underlying channel statistics. In other words, to comprehend the MDP dynamics, the UV agent requires specific assumptions about the statistical evolution of communication channels over time and space \cite{TSP2022Evmorfos, Diamantaras-MLSP-2019Optimal}.}

\subsubsection{DRL Algorithms}
DRL is built on a core theoretical framework, but there are many specific algorithms which handle the various cases discussed above.
\textcolor{black}{Table \ref{tab:deep_rl_algos} summarizes the characteristics of various model-free DRL algorithms.} Applications of these methods to optimize UV operation in communication systems are found in Section~\ref{sec:formulations}.
\begin{table}[h]
  \centering
\caption{The characteristics of various model-free DRL approaches.}
  \begin{tabular}{|c||c|c|c| c|}
    \hline
    \textbf{Algorithms} & \textbf{On-Policy} & \textbf{Off-Policy} & \textbf{Critic} &
    \hspace{-2pt}\textbf{Policy}\hspace{-2pt}\\
    \hline
    DQL \cite{mnih-Nature-2015human} &  & $\checkmark$ & $\checkmark$ & \\
    \hline
    Dueling DQL \cite{wang-ICML-2016dueling} &  & $\checkmark$ & $\checkmark$ & \\
    \hline
    SAC \cite{haarnoja-ICML-2018SAC}&  & $\checkmark$ & $\checkmark$ & $\checkmark$ \\
    \hline
    TD3 \cite{fujimoto2018addressing} &  & $\checkmark$ & $\checkmark$ & $\checkmark$ \\
    \hline
    PPO \cite{schulman-Arxiv-2017proximal}& $\checkmark$  &  & $\checkmark$ & $\checkmark$ \\
    \hline
     TRPO \cite{schulman2015trust}& $\checkmark$  &  & $\checkmark$ & $\checkmark$ \\
    \hline
    DDPG\cite{lillicrap-Arxiv-2015continuous}&   & $\checkmark$ & $\checkmark$ & $\checkmark$ \\
    \hline
     BPG\cite{bertsekas-AthenaScien-2019reinforcement, sutton-MITPRESS-2018reinforcement}& $\checkmark$  &  &  & $\checkmark$ \\
    \hline
   Zeroth-order DPG\cite{kumar2020zeroth}& $\checkmark$  &  &  & $\checkmark$ \\
    \hline
  \end{tabular}

  \label{tab:deep_rl_algos}
\end{table}

\subsubsection{Deep Learning}
Although the combination of deep learning and reinforcement learning is the most prominent paradigm for optimizing UV operation in wireless communications, there are other developed approaches that employ deep learning and do not consider a long time horizon of reasoning. Typically, these approaches are supervised learning methods that view the problem as autoregressive estimation. The work of \cite{9199789} introduces an IoT architecture involving multiple aerial UVs that facilitate communication among themselves and with ground base stations. In this architecture, one UV is designated as the leader, responsible for performing multiple beamforming. Optimizing the beamforming weights necessitates estimating the pose and position of the UVs. To address this challenge, the authors propose an autoregressive deep learning method that utilizes recurrent units and autoencoders for accurate pose and trajectory estimation. The work of \cite{9743298} considers a 3D communication scenario with IRS and UVs. The authors propose a deep learning regiment for channel tracking that consists of a pretrained neural network for denoising and a recurrent module that takes into account channel history. Finally, \cite{9296324} considers the motion of UVs in beamforming optimization and proposes a deep learning approach for beam alignment that accounts for UV mobility. The deep learning scheme comprises multiple recurrent units that predict the future position of the UV.

\section{Representative Problem Formulations}\label{sec:formulations}
As made apparent in previous sections, the set of possible problem formulations that arise when optimizing the operation of UVs in next-generation networks is incredibly large and diverse. This diversity stems not only from the myriad roles and applications of UVs in the network, as discussed in Section~\ref{sec:intro}, but also from the range of modeling choices, technical assumptions, and optimization tools that may apply to a single problem, as discussed in Sections~\ref{sec:modeling} and \ref{sec:approaches}.

Despite this variety, work in this area naturally clusters into a set of core problems, which we illustrate next with a series of representative problem formulations. These are presented with a level of abstraction so as to be agnostic to, \textit{e.g.}, UV kinematics or channel modeling, while highlighting the fundamental question posed in each. This abstraction is crucial in providing a unifying perspective for this complex landscape. Throughout this section, we will further put the challenges of 6G into perspective by highlighting what changes when considering next-generation communication systems. We first present analytic optimization formulations (see Section \ref{sec:approaches}), then present a selection of these in the context of reinforcement learning in Section~\ref{sec:rl_formulations}. Throughout, we discuss the importance of each problem and give examples from the literature.

\begin{table*}[]
    \centering
    \caption{Sample papers addressing the optimization of UVs in the role of providers.}
    \def\arraystretch{1.4}
    \begin{tabular}{|p{1.1in}|p{0.4in}|p{1in}|p{1in}|p{2.6in}|}
        \hlineB{5}
         \textbf{UV Mission}& \textbf{Ref.} & \textbf{Objective} & \textbf{Tools} & \textbf{Notes}\\
        \hlineB{4}
        \multirowcell{16}{\large UV as BS} & \cite{WCL2022Shamsabadi} & Worst-case SINR & MIP, SCA & OFDMA, sub-channel assignment, HAPS and TBSs.  \\\cline{2-5}
        & \cite{TWC2023Fang}  & Sum-rate & SCA & PD-NOMA, maritime environment.\\\cline{2-5}
        & \cite{ICC2016Bor-Yaliniz} & Number of users & MINLP & OMA, 3D placement. \\\cline{2-5}
        & \cite{DroneComm2022Ding}  & Sum-rate & SCA, Semi-definite relaxation & MIMO, full-duplex communication.\\\cline{2-5}
        & \cite{TWC2019Mozaffari}  & User-avg. latency & Optimal Transport & OFDMA, non-terrestrial network (NTN).\\\cline{2-5}
        & \cite{TWC2018Wu}  & Worst-case Throughput &  BCD/SCA & TDMA, trajectory planning.\\\cline{2-5}
        & \cite{bayerlein-IEEESPAWC-2018trajectory}  & Total communication rate &  RL & Single-input, multiple output (SIMO), Q-learning for motion control to maximize communication rate. \\\cline{2-5}
        & \cite{qin-TVT-2021distributed}  & Throughput &  DRL & MIMO, DPG for weighted throughput maximization subject to fairness.\\\cline{2-5}
        & 
         \cite{zhan-IEEETWC-2022energy}  & Energy consumption &  DRL & Motion Control of UV base stations to minimize energy expenditure subject to a constraint of perpetual connectivity.\\\cline{2-5}
        & 
        \cite{zeng-IEEEGLOBECOM-2019path}  & Task Completion  &  TD-learning & TD-learning with tile coding for motion control of UV base stations.  \\\cline{2-5}
        & 
         \cite{yin-IEEETVT-2019intelligent}  & Sum-rate  & DRL & UV continuous motion control for collective sum-rate maximization.
        \\\cline{2-5}
        & \cite{mohammadi-Arxiv-2022analysis}  & Throughput &  DRL & MIMO, SARSA and offline RL for throughput maximization under fading.\\\hlineB{4}

        \multirowcell{3}{\large ISAC} & \cite{TxITS2023Li} & Generic sensing and communication & Calculus of variations & Sensing performance measured with sensing SNR.\\\cline{2-5}
        & \cite{TWC2023Meng} & Sum-rate & MIP, BCD, SCA & MIMO, SDMA for sensing and communication.\\\hlineB{4}

        \multirowcell{25}{\large Cooperative\\\large Relay}  & \vspace{-0.08in}\hspace{-0.1in}\makecell{\cite{Kalogerias-IEEETSP-2018Spatially},\\\cite{Kalogerias-IEEEICASSP-2016mobile}} & SINR & Distributed stochastic optimization & Distributed beamforming. A queue of previous positions and channels is used to solve a stochastic surrogate for channel prediction.\\\cline{2-5}
        & \cite{TRo2012Yan} & Bit-error probability, motion energy & KKT conditions & Multihop robotic relay, uncertain channel environment with obstacles, Gaussian-process based channel prediction.\\\cline{2-5}
        &\cite{Access2013Kumar} & Abstract task completion & SOCP, control barrier functions, $\text{RRT}^*$ & Multihop relay. Extensive real-world experimentation.\\\cline{2-5}
        & 
        \vspace{-0.08in}\hspace{-0.1in}\makecell{\cite{TCNS2017Muralidharan},\\\cite{ICASSP16_MdharanMostofi}} & Motion and communication energy & Multiple-choice knapsack problems &  Distributed robotic beamforming, Accounts for uncertainty in shadowing and multipath fading via stochastic channel prediction.\\\cline{2-5}
        & \vspace{0.0in}\hspace{-0.1in}\makecell{\cite{TSP2022Evmorfos},\\ \cite{ICASSP2022Evmorfos}} & SINR & DRL & Distributed beamforming relay. Fourier features as preprocessing to estimate the value function depending on spatio-temporally correlated channels.\\\cline{2-5}
        & \cite{kalogerias2014mobi} & Secrecy rate & Stochastic programming & Distributed beamforming relay. Distributed stochastic programming for secrecy rate maximization under cooperation.\\\cline{2-5}
        & 
        \cite{hassan-IEEEICC-20223to} & Throughput & DRL & Single UV relay. Motion control for average system throughput maximization.\\\cline{2-5} 
        
        & \cite{mozaffari2015drone} & Transmit power & BCD & BCD to optimize UV positioning in multihop relaying.\\\cline{2-5}
        & \cite{farooq2018multi} & Maintaining connectivity & Particle Swarm Optimization & Particle Swarm Optimization for UV placement for multihop relaying.\\\cline{2-5}
        & \cite{kang20203d} &  Min communication rate & BCD, Gibbs sampling & 3D UV placement and power and bandwidth allocation for multihop relaying.\\\cline{2-5}
        
        & \cite{shin-IEEEWCL-2023sub} & Energy & Multi-agent DRL & Single UV relay. Multi-agent RL surpasses hybrid precoding for energy minimization.\\
         \hlineB{5}
    \end{tabular}
    \label{tab:provider_summary}
\end{table*}

\subsection{Unmanned Vehicles as Providers}
 In this subsection, we present three broad categories of problems in which UVs act as providers. In the first, UVs provide connectivity to the core network for multiple users, possibly in environments which include terrestrial BS (TBS) as well. In the second, the UV uses ISAC to perform a sensing task while simultaneously communicating. In the third, UVs work cooperatively to extend connectivity by relaying either with multihop communication or distributed beamforming.

\subsubsection{\textbf{Providing Multiple Access with Resource Allocation}}
Extending networks to underserved, isolated, or simply poorly-connected areas through the use of UVs is an exciting proposition which has received significant attention in both academia and industry. More generally,  when UVs act as base stations (UV-BSs), either as a cost-effective solution to closing the digital divide, as providers of connectivity for isolated or poorly-connected areas, or as dynamic assets to relieve temporary high-traffic loads in urban areas, a core problem is resource allocation: What is the best way to distribute a UV-BS's finite wireless resources (power, time, bandwidth, and spatial degrees of freedom) across multiple concurrent users? 

\begin{figure}
    \centering
    \includegraphics[width=3.4in, trim={2in, 2.1in, 4.5in, 1.3in}, clip]{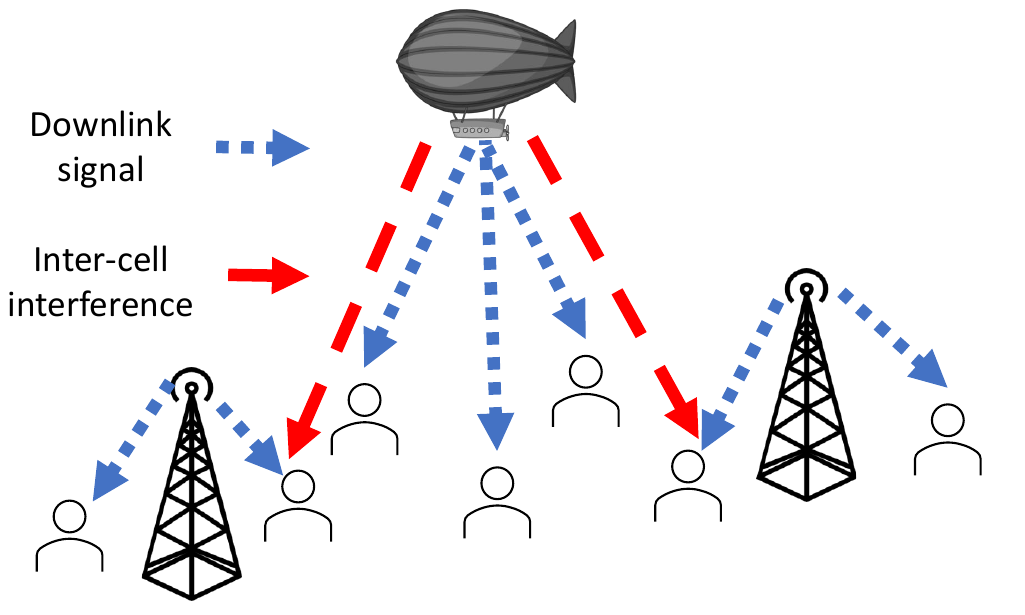}
    \vspace{-0.1in}
    \caption{A UV acting as a base station must support concurrent users while mitigating inter-cell interference.}
    \label{fig:multi_access}
    \vspace{-0.2in}
\end{figure}

Resource allocation over multiple simultaneous links for UV-BS differs from existing results on multi-access for TBS in two important ways. First, the mobility of UV-BSs creates an additional degree of freedom for planning. Second, the elevated position of aerial UV-BSs results in a high probability of strong LOS channels to unintended receivers, so that mitigating inter-cell interference takes on greater importance. A high-level depiction of the problem setting is shown in Fig.~\ref{fig:multi_access}.

To pose a general problem, let $\mathcal{U}$ denote the set of users, $\mathcal{C}_i, i \in \mathcal{U}$ indicate a per-user communication metric (Section~\ref{sec:modeling}), and $f_\text{comm}$ here be an abstract aggregation operator which takes as input each user's link-level performance and produces a single, system-level metric. Let $\mathcal{B}_{\text{UV}}$ denote the set of UV-BS, $\mathcal{B}_{\text{TBS}}$ the set of TBSs, with $\mathcal{B}=\mathcal{B}_{\text{UV}} \cup \mathcal{B}_{\text{TBS}}$. The problem consists of finding the binary BS-to-user associations, $\{a_{bi}\}_{(bi)\in\mathcal{B}\times \mathcal{U}}$, the locations of the UV BS, $\{q_b\}_{b\in \mathcal{B}_{\text{UV}}}$, and the resources allocated from each BS to each user, $\{\mathcal{R}_{bi}\}_{(bi)\in\mathcal{B}\times \mathcal{U}}$, which maximize the aggregate metric of system performance, subject to positioning, communication resource, and QoS constraints. The problem is given below:
\begin{maxi!}[1]
    {\substack{\{a_{bi}\},\\ \{q_b\},\{\mathcal{R}_{bi}\}}}                              
    {\hspace{-0.2in}f_\text{com}\bigg(\{ \mathcal{C}_{i} \big(\{a_{bi}\},\, \{\mathcal{R}_{bi}\},\, \mathcal{H}_i (\{q_b\})\big)\,\}_{i \in \mathcal{U}}\bigg)\label{objective:RA_multi}}   
    {\label{opt:RA_multi} }             
    {}                                
    \addConstraint{g_{\text{com}}(\{\mathcal{R}_{bi}\}) \leq 0,\,\forall b\in \mathcal{B} \label{constraint:RA_multi.resrouce_constraint}}
    \addConstraint{\tilde{\mathcal{C}}_i\big(\{a_{bi}\},\, \{\mathcal{R}_{bi}\}, \mathcal{H}_i(\{q_b\}) \big) \geq \tilde{\mathcal{C}}_{i,\text{req}},\,\forall i \in \mathcal{U}\label{constraint:RA_multi.performance_constraint}}
    \addConstraint{\sum_{b \in \mathcal{B}} a_{bi} = 1,\,\forall i \in \mathcal{U}}
    \addConstraint{a_{bi} \in \{0,\,1\},\,\forall\, b\in\mathcal{B},\,i\in\mathcal{U}}
    \addConstraint{q_b \in \mathcal{Q}_b,\, \forall b\in\mathcal{B}, \label{constraint:RA_multi.position_constraint}}
\end{maxi!}
where, for brevity, we have written the association and resource allocation variables as $\{z_{bi}\}$ rather than $\{\{z_{bi}\}\}_{(bi)\in\mathcal{B}\times \mathcal{U}}$ for $z\in \{a, \mathcal{R}\}$, and $\mathcal{H}_i(\{q_b\})$ represents the set of abstract channels from all BS to user $i$ (see Section~\ref{sec:modeling}). The link-level metric $\mathcal{C}_i$ is a function of BS-to-user associations as well as resource allocation. and the communication channels as determined by the BS positions. Constraint (\ref{constraint:RA_multi.performance_constraint}) is a QoS constraint which requires that a link-level metric $\tilde{\mathcal{C}}_i$ exceeds a value $\tilde{\mathcal{C}}_{i,\text{req}}$ for each user. Constraint (\ref{constraint:RA_multi.resrouce_constraint}) ensures communication resources are not over-allocated, and (\ref{constraint:RA_multi.position_constraint}) captures positioning constraints such as obstacles, no-fly zones or fixed flying altitudes. 

There are several important challenges raised in this generic problem statement. The binary user association variables introduce a combinatorial element to the problem, which is naturally expressed as a MIP. Non-convexity enters through inter-cell interference, highlighted by the dependence of $\mathcal{C}_u$ and $\tilde{\mathcal{C}}_u$ on \textit{all} resource allocation variables, as well as through non-convex spatial variations in the communication channels, which enter the problem through $q_b$.

Many problems may be formulated through specific choices of the link-level metric, aggregation operator, resources allocation constraints, and positioning constraints. We next review a few interesting examples.

\textbf{Providing Multiple Access with Given Position:} In some cases, such as when dealing with quasi-stationary HAPSs, the positioning of the UV is assumed fixed, so that optimization only occurs over the communication resource variables. For example, in \cite{WCL2022Shamsabadi}, a system consisting of multiple TBSs and a single HAPS serve a set of ground users using OFDMA. The authors formulate a problem which seeks to maximize the minimum SNR over all users by optimally selecting a sub-carrier frequency channel and power allocation for each user. An iterative procedure based on  SCA repeatedly solves a MIP to find solutions to the problem which enhance the worst-case SNR among the users.

A multi-UV variation appears in \cite{TWC2023Fang}, in which a team of UAV-BSs, TBSs, and a satellite jointly provide communication to several maritime users using non-orthogonal multi-access (NOMA). The transmit powers of the UAV-BSs and TBSs are jointly optimized to maximize the sum-rate under the constraint that interference does not significantly degrade communication for the users connected via satellite. To solve the problem, an iterative SCA method is proposed, and results show that the NOMA multi-access scheme greatly improve the sum-rate compared to orthogonal multi-access (OMA) schemes in this maritime scenario.

\textbf{Multiple Access with Optimized Positioning:} In many cases, a UV-BS deployed to service an area will operate from or near a fixed point. In \cite{ICC2016Bor-Yaliniz}, the authors examine the 3D-placement problem for a UAV which is deployed to improve the network capacity. The UAV services all users for which the channel between the user and the UAV achieves some threshold value, and the 3D placement of the UAV must be chosen to maximize the total number of users serviced by the UAV. The problem is formulated as a mixed-integer nonlinear program (MINLP) which is solved using an interior-point method and a bisection search. 

In addition to UAV placement, \cite{DroneComm2022Ding} optimizes beamforming weights and uplink-user transmit power in a scenario where a full-duplex UAV-BS simultaneously handles uplink and downlink traffic from distinct sets of users with LOS/NLOS mmWave channels. The challenging non-convex problem is solved by first optimizing the UAV's location using SCA, then finding the UAV-BS beamforming weights and uplink user transmit powers by relaxing a rank-1 matrix constraint. It is shown that omitting optimization of the UAV-BS's location reduces sum-rate by nearly $1\,$b/s/Hz, and that greater performance loss occurs when using only half-duplex.

The work in \cite{TWC2019Mozaffari} develops a fully aerial network, with designated UAV-BSs providing connectivity to UAV end users using OFDMA. The authors pose the problem of partitioning the coverage region among the UAV-BS to achieve the minimal average communication latency, and the problem is solved using optimal transport \cite{OptTrans2008Villani}.

\textbf{Multiple Access with Trajectory Optimization:} When optimizing over time, additional kinematic and energy constraints may enter the problem. We express these generically as 
\begin{equation}\label{constraint:generic_kinematics}
    \begin{rcases}
        \hspace{-0.2in}g_{\mathcal{K}}(\{q[k]\}_{k=1}^K)& \leq 0\\
        \hspace{-0.2in}h_{\mathcal{K}}(\{q[k]\}_{k=1}^K)& = 0
        \end{rcases}\,\text{Motion constraints},
\end{equation}
where $k$ and $K$ are the discrete-time index and horizon, respectively. For specific examples of such constraints, see Section~\ref{sec:modeling}. For brevity, in the sequel we denote $\{q[k]\}_{k=1}^K$ with $\{q[k]\}$, unless clarity demands otherwise.

The authors of \cite{TWC2018Wu} consider multiple UAV-BSs moving from given initial locations to given final locations while providing connectivity to a set of end users using TDMA. A problem is formulated which maximizes the minimal time-averaged throughput among end users by jointly optimizing trajectories, transmit powers, and user associations. To solve the problem, BCD is used in conjunction with SCA to find solutions. Numerical examples show that utilizing mobility significantly increases the minimal throughput compared to the placement of static BSs.

\textbf{Takeaways:}
Thus far, we have looked at resource allocation and positioning for UV-BSs supporting multiple users. These problems are characterized by binary user-BS association variables and a need to mitigate inter-cell interference. Several examples have illustrated the performance gains achievable by deploying UVs and exploiting their mobility, and SCA has proved to be a useful tool for solving these problems. Emerging 6G multi-access techniques, such as PD-NOMA, RSMA, and OFTS, introduce new variations in user-to-resource association and interference modeling. For example, re-use of PRBs may eliminate constraints tying each PRB to at most one user, while simultaneously introducing greater complexity in calculating interference.

We next consider ISAC, in which shared time, frequency, and power resources are allocated not only among different communication users but also more broadly between sensing and communication tasks.

\subsubsection{\textbf{Integrated Sensing and Communication (ISAC)}}
In 6G networks, the high frequencies that will enhance communication capacity also enable sensing applications, leading 3GPP to designate ISAC as a new core usage scenario. Fully integrating these two applications requires combining hardware, signal processing pipelines, and waveforms, as well as understanding how to allocate limited resources between these two tasks. A thorough survey of the area may be found in \cite{liu2020joint,zhang2021overview,JSAC2022Liu}.

\begin{figure}
    \centering
    \includegraphics[width=3.4in, trim={0.5in, 2.55in, 3.5in, 1in}, clip]{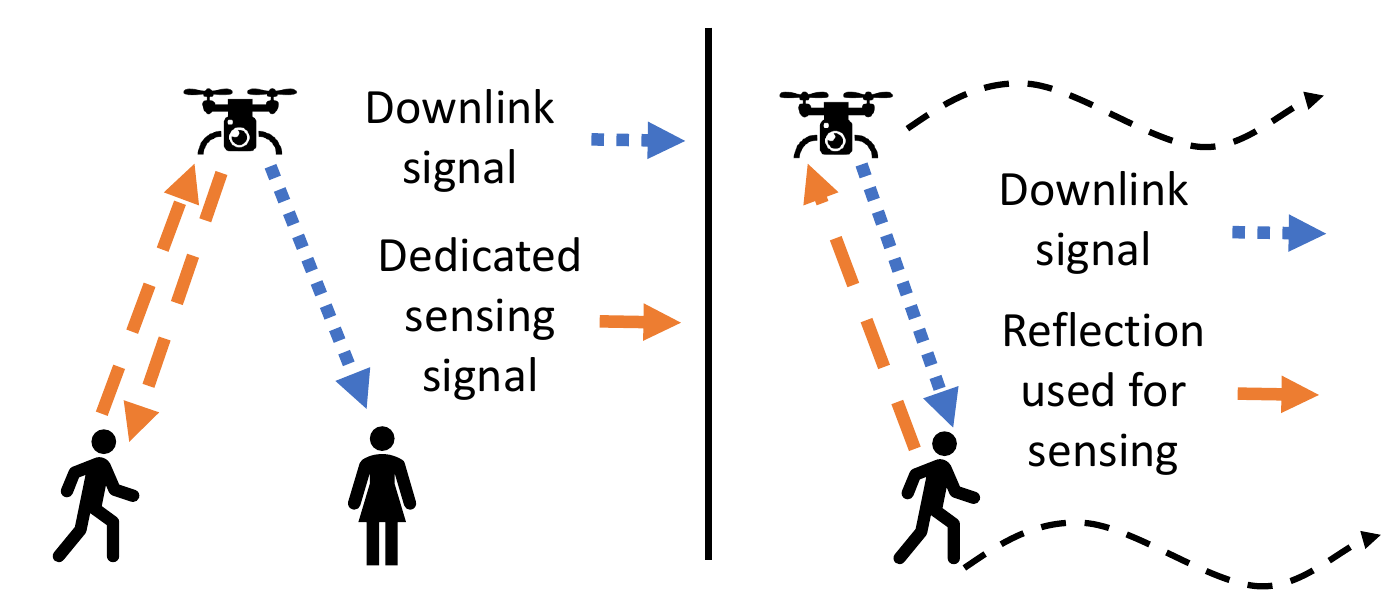}
    \caption{ISAC may use either orthogonal sensing and communication signals, or a single signal may be used for both applications.}
    \label{fig:alt_ISAC}
\end{figure}

The use of UVs in the context of ISAC has garnered significant attention, even prior to the advent of 6G networks \cite{Depatla2017APM}, and many advantages of using UVs for communication extend to sensing-related tasks. Mobility permits repositioning to find better sensing locations, while the high probability of LOS paths for UAVs/HAPSs avoids occlusion in the sensing domain. 
\cite{WC2023Meng} and \cite{CommsMag2023Mu} provide additional discussion of ISAC specifically in the context of UAVs.

To pose a general problem, we consider two sets of resource-allocation variables, $\mathcal{R}_{C}$ and $\mathcal{R}_{S}$, representing the resources allocated to communication and sensing, respectively. While current sensing standards treat these resources as orthogonal, as in 802.11bf which uses time-division to isolate sensing and communication signals \cite{CommsMag2023Chen}, finding dual-function waveforms that achieve sensing and communication simultaneously is an area of active research. For a single UV, a general problem formulation is given below:
\begin{maxi!}|s|[2]
 {\substack{\{q[k]\}, \{\mathcal{R}_{C}[k]\},\\ \{\mathcal{R}_{S}[k]\}}}                              
{f_{\text{GLB}}(f_{\text{com}},\, f_{\text{sense}})  \label{objective:ISAC}}   
{\label{opt:ISAC} }             
{}   
\addConstraint{(\ref{constraint:generic_kinematics})\;\text{(Motion constraints)}\notag}
\addConstraint{g_{\text{com}}(\mathcal{R}_{C}[k],\,\mathcal{R}_{S}[k]) \leq 0,\,\forall K\label{constraint:ISAC.shared_resrouces}}.
\end{maxi!}

The objective in (\ref{objective:ISAC}) captures the trade-off between communication performance, $f_{\text{co}}$, and sensing performance, $f_{\text{sense}}$, and both are functions of all optimization variables. Either the sensing or communication performance could alternatively be treated as a constraint, and Constraint~(\ref{constraint:ISAC.shared_resrouces}) captures the fact that resources are shared across sensing and communication applications. We next provide two interesting examples.

In \cite{TxITS2023Li}, the authors study the optimal trajectories for a UAV tasked with sensing and communicating with a mobile ground node. They show that the sensing and communication quality at different points can be viewed as an artificial potential field, so that maximizing an abstract communication and sensing service can be found by determining the shape of a catenary \cite{OE2005Griva}. Assuming path loss-only communication models, analysis shows the global optimality, existence, and uniqueness of the given solution.

The sensing performance may be taken as a constraint rather than part of the objective. For example, the work in \cite{TWC2023Meng} investigates a multi-antenna UAV which communicates with multiple users using TDMA while also periodically sensing a number of targets. To optimize the total communication rate, communication user and sensing targets are scheduled along with corresponding beamforming weights, subject to the constraint that when sensing is scheduled, the sensing SNR (distinct from the communication SNR) must exceed a given threshold. The problem is solved using a general BCD approach, and results show a trade-off between achievable communication performance and sensing frequency.

\begin{figure}
    \centering
    \includegraphics[width=3.4in, trim={3in, 0.2in, 2.5in, 0.5in}, clip]{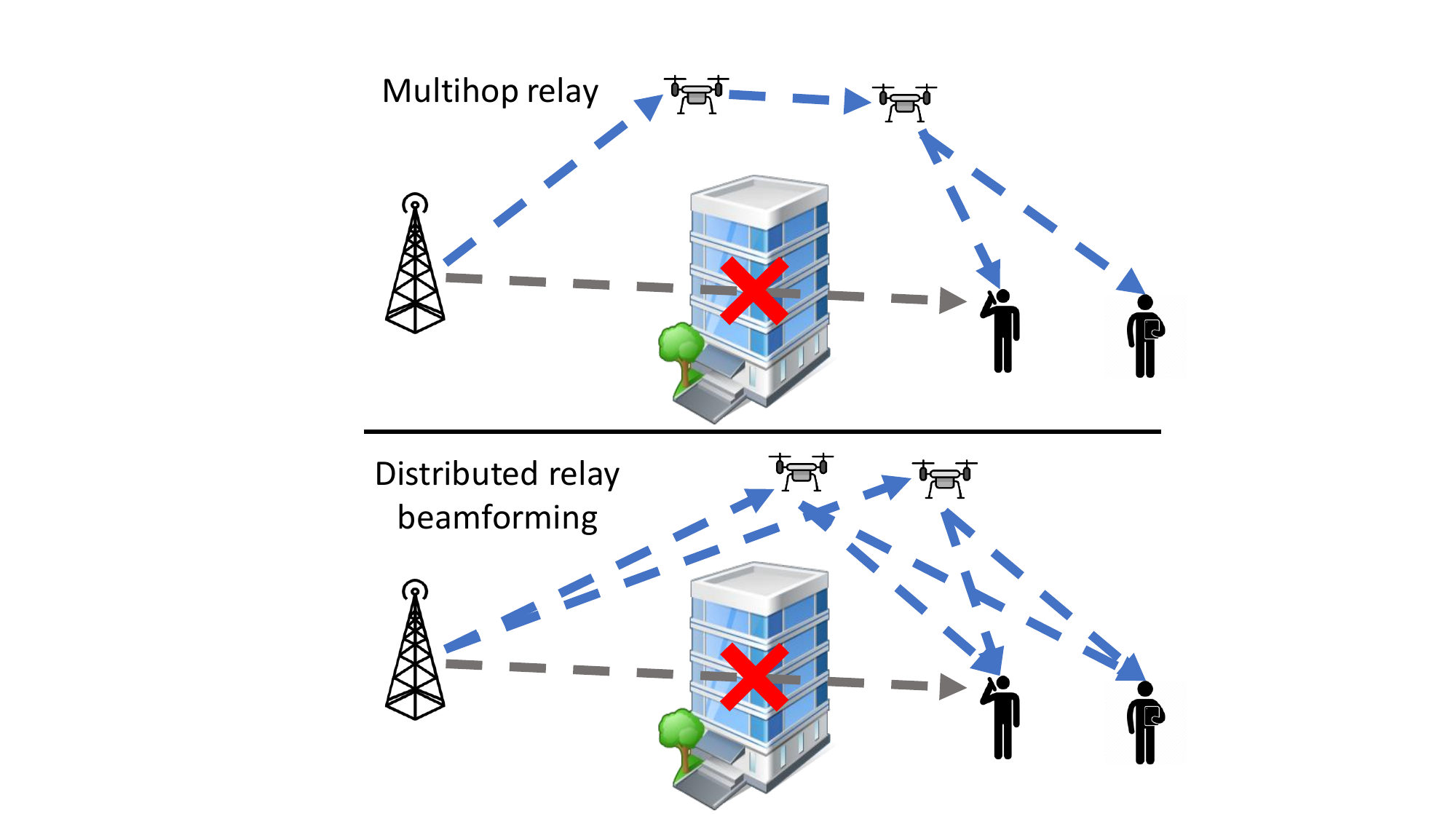}
    \caption{UV-enabled distributed relay beamforming can extend existing terrestrial networks.}
    \label{fig:relay}
    \vspace{-0.2in}
\end{figure}

\textbf{Takeaways:}
While many capabilities of 6G networks are improvements over similar 5G capabilities, ISAC is a novel emerging paradigm. Similar to communication, UVs can use mobility to improve sensing in ISAC scenarios, but UVs must balance resource allocation and trajectory planning between communication and sensing objectives, which may either be competing or complementary. As the technology matures, the place of UVs in enabling these applications will continue to develop\cite{WC2023Meng}.

\subsubsection{\textbf{Relaying and Cooperative Communication}}
In many scenarios, UVs can work cooperatively to enhance and extend communication networks. A key example is the use of UVs as relays, either by creating a chain of communication links to extend connectivity via multihop relay or coordinating transmission to achieve distributed beamforming.

A generic setting is the following: $M$ sources aim to communicate with $M$ destination nodes, but direct communication is not possible due to obstruction of a LOS path or significant distance. Facilitating this communication are $N$ UV relays which will move to locations $q_{n},\,n\in\{1,...,N\}$. The communication performance for each source-destination pair (SDP) is given by $C_m,\,m\in\{1,...,M\}$, which represents metrics such as data rate or BER. For each SDP, let $S_m$ be an abstract object representing the routing path from source to destination (e.g., a sequence of UVs), and let $\mathcal{R}_{n,m}$  denote the communication resources, allocated by UV $n$ to SDP $m$, with $\mathcal{R}_m=\{\mathcal{R}_{n,m}| \, n=1,..., N\}$. Then, a high-level problem can be formulated as below:

\begin{maxi!}|s|[2]
    {\substack{\{q_{i}\}_{i=1}^{N}, \{\mathcal{R}_{m}\}_{m=1}^M,\\ \{S_m\}_{m=1}^M}}                              
    {\sum_{m=1}^{M} C_m(\mathcal{R}_m, S_m, \{q_i\})}   
    {\label{opt:UV_relay} }             
    {}                                
    \addConstraint{\{q_{i}\}_{i=1}^{N}\in \mathcal{Q}\label{constraint:UV_relay.placement_constraint}}
    \addConstraint{g_{\text{com}}(\{\mathcal{R}_m\}) \leq 0 \label{constraint:UV_relay.comm_constraints}}
    \addConstraint{g_r(\{S_m\}) \leq 0\label{constraint:UV_relay.route_constraints}} .
\end{maxi!}
Constraint (\ref{constraint:UV_relay.placement_constraint}) ensures that the placement of the UVs is valid, e.g., enforcing minimum distances between UVs, while Constraint (\ref{constraint:UV_relay.comm_constraints}) ensures that communication resources are not over allocated. Constraint (\ref{constraint:UV_relay.route_constraints}) requires that communication routing is valid, e.g., there are no cycles. For clarity of exposition, we have written the SDP-level communication performance, $C_m$, as a function of only the resources allocated to SDP $m$, but more generally, e.g., when communication for the different SDPs does not occur over orthogonal channels, $C_m$ may be a function of resources allocated to other SDPs as well. We next look at examples of these types of problems from the literature, focusing first on multihop routing, then proceeding to distributed beamforming approaches.

\textbf{Multihop mobile relaying:} 
\textcolor{black}{Multihop relaying \cite{zafar2012analysis} (see Fig. \ref{fig:relay} (top)), departs from the conventional point-to-point communication model by introducing a chain of data transmission through intermediary nodes. Such approaches can extend coverage and hedge against NLOS conditions. UVs can effectively serve as relays in multihop scenarios, as their inherent mobility and adaptability prove beneficial in responding to evolving communication conditions, such as variations in communication channels and the mobility of end-users.}

As an example of such a scenario, \cite{TRo2012Yan} considers multiple UGVs moving from initial positions to find locations which minimize the end-to-end BER for a multihop relay scenario in the presences of obstacles and stochastic communication channels. For certain conditions, the problem is shown to be concave, and the KKT conditions can be used to produce solutions. \cite{Access2013Kumar} considers an ad hoc network of robots which must support multiple source-destination communication flows while performing an abstract task. For tractability, the problem is decomposed, with communication routing found using a second-order cone program (SOCP) \cite{Convex2004Boyd}, local control laws derived using barrier functions \cite{ECC2019Ames}, and $\text{RRT}^*$ \cite{IJRR2011Karaman} providing high-level trajectories.

\textcolor{black}{In \cite{mozaffari2015drone}, the authors consider a problem such as the one described above. They propose a BCD \cite{wright2015coordinate} method to find the positions of the UVs which minimize the total transmit power under a coverage constraint. The work of \cite{farooq2018multi} examines a similar UV multihop scenario and proposes a particle swarm optimization approach to design the UV positions for maintaining connectivity. Finally, the authors of \cite{kang20203d} examine a scenario with multiple source-destination pairs where communication is facilitated by multiple UV relays in a multihop setting. They propose an alternating optimization scheme that combines BCD with Gibbs-sampling \cite{gelfand2000gibbs} to jointly optimize the UV positions as well as the bandwidth and power allocation, with the objective of maximizing the minimum communication rate among source-destination pairs.}

\textcolor{black}{\textbf{Distributed relay beamforming}: Distributed relay beamforming \cite{zheng2009collaborative}, akin to multihop relaying, involves source-destination pairs seeking information exchange. Relays between these pairs receive signals simultaneously and transmit them collectively to destinations in synchronous fashion. Despite its classification as a multihop relaying case with two hops, it is often studied independently due to nuanced domain-specific considerations. These nuances involve cooperative relay decision-making to form a unified beam at the destination.} 

In recent years, beamforming in the context of UVs has gained significant attention. For example, \cite{ICASSP16_MdharanMostofi,TCNS2017Muralidharan}  focus on minimizing both motion and communication energy while jointly optimizing the positions of multiple UGVs and their corresponding beam weights for distributed robotic beamforming. The paper considers an uncertain channel environment and shows how to incorporate channel prediction using a Gaussian-process-based prediction model. Near-optimal solutions are then found through a series of multiple-choice knapsack problems \cite{PISINGER1995394}, which can be solved efficiently despite being NP-hard in the general case.

\textcolor{black}{The study in \cite{koyuncu2008distributed} explores scenarios where multiple relays engage in distributed beamforming for a single SDP, incorporating quantized feedback. The authors recommend using the Generalized Lloyd Algorithm \cite{sabin1986global} to design the quantizer for specified beamforming weights. In a related domain, \cite{havary2008distributed} delves into employing multiple relays for distributed relay beamforming in a two-transceiver system. Two scenarios are considered: optimizing relay beamforming weights to minimize total transmit power while meeting receiver SNR constraints, and maximizing the lowest SNR between two receivers under a total transmit power constraint. Proven unique solutions and proposed iterative schemes showcase linear complexity per iteration.
In cooperative jamming, \cite{kalogerias2013mobile} focuses on UV motion control and noise nulling to maximize collective secrecy rates. \cite{kalogerias2014mobi} optimizes relay positioning for acceptable receiver-side QoS and maximizes ergodic secrecy rates with an eavesdropper.}

\textcolor{black}{The works \cite{Kalogerias-IEEEICASSP-2016mobile, Kalogerias-IEEETSP-2018Spatially} consider scenarios with one SDP and multiple UV relays, aiming to select UV positions and beamforming weights to maximize a utility metric. Assuming underlying channels evolve as Gaussian processes with correlations across time and space, the authors propose a distributed stochastic optimization method. This approach predicts channels for candidate UV positions based on historical data, achieving closed-form expressions for optimal beamforming weights once positions are determined.}

\textbf{Takeaways:} The use of UVs as mobile relay nodes introduces the element of dynamic positioning, an additional controllable parameter for optimizing various aspects of system performance. When multiple UVs provide this relay service together, their operation must be optimized jointly. The cooperative paradigm, whether the team of UVs acts as a single relay chain or employs distributed beamforming, greatly influences optimal positioning and resource allocation. 

Technologies used to realize the 6G vision will introduce new challenges and use cases for UV relays. With the acute penetration loss which accompanies higher communication frequencies, relaying may become a more important paradigm, particularly in highly cluttered environments where direct LOS paths between a transmitter and receiver are less likely. Co-deployment with RIS, which can act as passive relays, will add another dimension to these optimization problems.


\begin{figure}
    \centering 
    \includegraphics[width=3.4in, trim={0.1in, 0.5in, 0.3in, 0.1in}, clip]{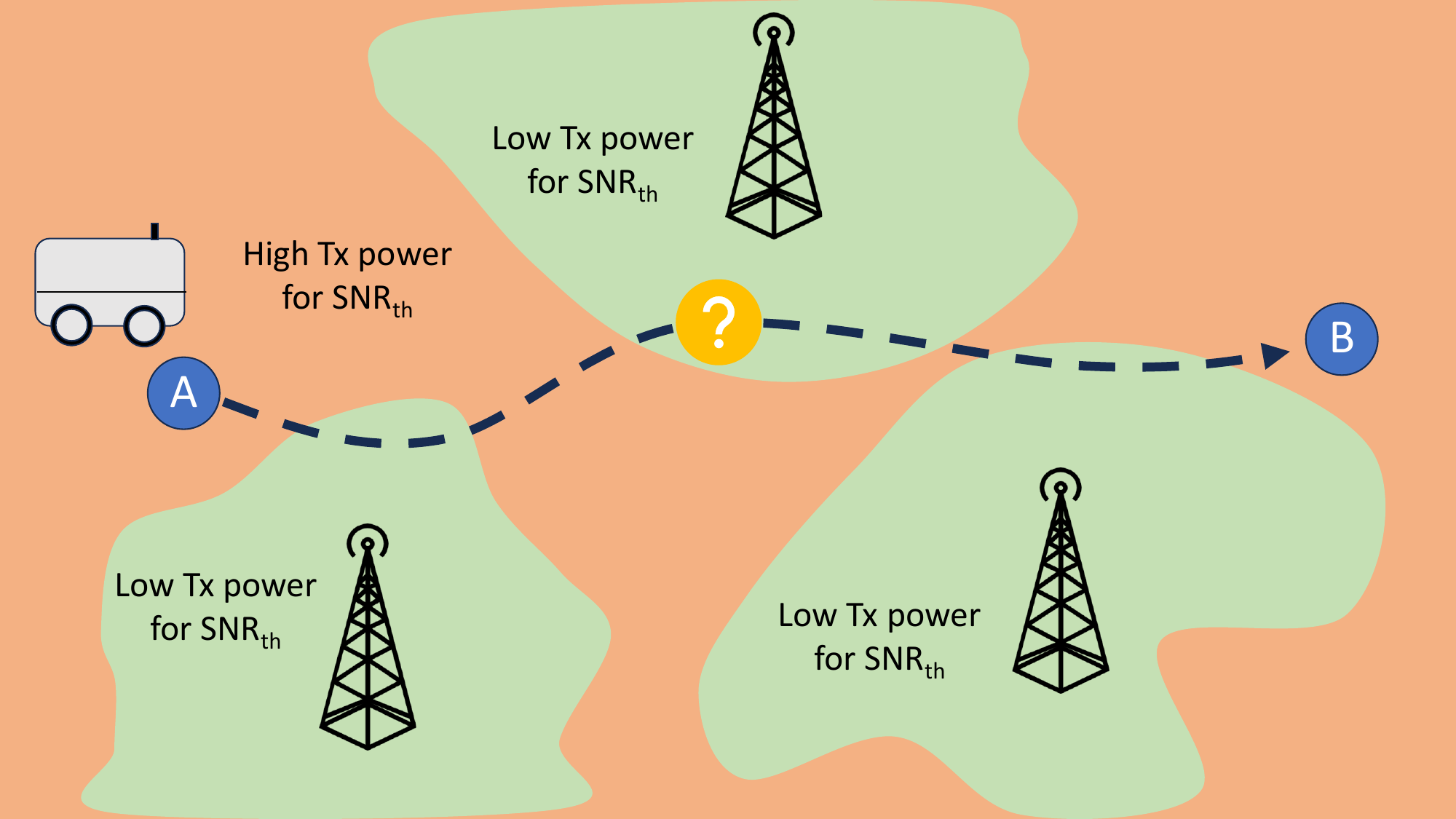}
    \caption{Maintaining connectivity: A fundamental problem for UVs is how to efficiently move between two locations while maintaining connectivity.}
    \label{fig:maintain_connectivity}
\end{figure}
\subsection{Unmanned Vehicles as Consumers}
Supporting connectivity for UVs is an expected feature of 6G networks, with related standards actively under development \cite{3GPPTR36777}. In this section, we present several fundamental problem formulations in which UVs consume the services provided by 6G systems, beginning with the problem of maintaining connectivity, then moving towards more complex problems of wireless data-offload scheduling with data causality constraints in the context of IoT and WSN data collection. 

\begin{table*}[]
    \centering
    \caption{Sample papers addressing the optimization of UVs in the role of consumers.}
    \def\arraystretch{1.2}
    \begin{tabular}{|p{1.1in}|p{0.3in}|p{1.1in}|p{1.3in}|p{2.3in}|}
        \hlineB{5}
          \textbf{UV Mission} & \textbf{Ref.} & \textbf{Objective} & \textbf{Tools} & \textbf{Notes}\\
        \hlineB{4}

          \multirowcell{5}{\large Maintaining\\\large Connectivity} & \cite{ICC2021Hurst} & Motion and communication energy & $\text{RRT}^*$ & Obstacle avoidance, uncertain channel environment. \\\cline{2-5}
         & \cite{TC2019Zhang} & Mission time & Graph theory, Convex optimization  & UAV associates with nearest BS. \\\cline{2-5}
         & \cite{ICC2021Cherif} & Handovers & Dynamic programming & Cargo UAV application. \\\hlineB{4}

         \multirowcell{2}{\large Reconnecting to\\\large the Network} & \cite{TR2019Muralidharan} & Distance to connectivity &  Game theory, Graph theory, Dynamic programming & Probability of success modeled as independent across discretized space. Shown to be NP-hard. \\\cline{2-5}
         & \cite{Autonomous19_MuralidharanMostofi} & Distance to connectivity & Ornstein-Uhlenbeck process & Accounts for log-normal shadowing and generic multipath fading. \\\hlineB{4}

         \multirowcell{4}{\large Upload\\\large Scheduling}  & \cite{TWC2013Yan} & Motion and communication energy &  KKT conditions & Velocity and transmission rate optimized over a given trajectory.\\\cline{2-5}
         & \vspace{-0.08in}\makecell{\hspace{0.02in}\cite{ACC2015Ali},\\\cite{TCNS2018AliCai}} & Motion and communication energy & Optimal control & Abstract Bolza optimal control problem.\\\hlineB{4}

         \multirowcell{20}{\large Site Visiting}  & \cite{nguyen-IEEETC-20223d} & Sum-rate and completion time & DRL & Deep Q-learning for motion control to optimize a convex combination of sum-rate and task completion time.\\\cline{2-5}
         & \cite{TWC2023Hurst} & Avg. Wait Time & Polling systems, MISOCP, stochastic planning & Data collection and relay. Optimal robotic routing locations/policies. Long-run analysis of stationary regime\\\cline{2-5}
         & \cite{CDC2023Hurst} & Avg. Wait Time & Polling systems, DRL & Data collection and relay. Dynamic visit sequence. Handles long-run average cost criterion.\\\cline{2-5}
         & \cite{Oubati-IEEETVC-2022Sync} & Data collection and wireless energy transfer & Multi-agent DRL& Split the UVs into two groups (data harvesting and energy transfer) and employ multi-agent DRL since the policies are correlated.\\\cline{2-5}
         & \cite{Li-IEEEIoTJ-2021Joint} & Data rate & DRL for partial observability & Buffer overflow can cause unsuccessful transmission that induces partial observability.\\\cline{2-5}
         & \cite{TCNS2014Yan} & Motion and communication energy & MILP & Sensing mission, collects set amount of data from each site.\\\cline{2-5}
         & \cite{TCNS2022Hong} & Motion and communication energy & Monte Carlo tree search & Trajectory must pass within a certain distance of each site.\\\cline{2-5}
         & \cite{TWC2017Mozaffari_IOT} & Device energy & Sequential \hspace{-0.05in} quadratic programming & Energy reduction for IoT device communication\\\cline{2-5}
         & \cite{TWC2020Samir} & Number of devices served & MINLP, SCA, Branch, reduce, \& bound & Shown to be NP-hard.\\\cline{2-5}
         & \cite{GLOBECOM2020Ghdiri} & UV energy & Vehicle routing problem & Capacitated vehicle routing problem with time windows.\\\cline{2-5}
         & \cite{ACMTSN2014Ghaffarkhah} & Motion and communication energy & MILP & Sites visited on Hamiltonian path.\\\cline{2-5}
         & \cite{TWC2023Gao} & Peak/Avg. AoI & Ant colony optimization & Sensor-node-to-UAV wireless communication, UV moves to BS before offloading (data muling).\\\cline{2-5}
         & \cite{TVT2019Abd-Elmagid} & Avg. AoI & BCD, SCA, Convex optimization & Considers communication over both WSN-to-UAV and UAV-to-BS links.\\         
         \hlineB{5}
    \end{tabular}
    \label{tab:consumer_summary}
    \vspace{-0.2in}
\end{table*}


\subsubsection{\textbf{Maintaining Connectivity while in Motion}}
Consider a UV which must move between two points while maintaining connectivity with a remote server in order to, \textit{e.g.}, receive control and non-payload communication (CNPC) \cite{KaB2012kerczewski} or upload sensed data in real time. These problems can be frequently posed as shortest-path problems, with distance metrics that account for transmission and motion power. As a baseline example, consider a UV moving from a given initial position, $q_0$, to a final position, $q_f$, while maintaining the SNR between the UV and a remote BS above a minimum threshold, $\text{SNR}_\text{th}$, as illustrated in Fig.~\ref{fig:maintain_connectivity}. This baseline formulated is

\begin{mini!}|s|[2]                   
    {\substack{\{q[k]\}_{k=1}^K,\\ \{\mathcal{R}[k]\}_{k=1}^K, K}}                
    {\hspace{-0.2in}\lambda_1 \mathcal{E}_{\mathcal{K}}(\{q[k]\}) + \lambda_2 \mathcal{E}_{\text{com}}(\{\mathcal{R}[k]\}) + \lambda_3 \Delta t K \label{objective:user.MC}}   
    {\label{problem:user.MC}}             
    {}                                
     \addConstraint{(\ref{constraint:generic_kinematics})\;\; \text{Motion constraints}\notag}
    \addConstraint{\hspace{-0.2in}g_{\text{com}}(\mathcal{R}[k])\leq 0 ,\,\forall k \label{constraint:user.MC.comm_power}}
    \addConstraint{\hspace{-0.2in}\text{SNR}(q[k], \mathcal{R}[k])\geq \text{SNR}_\text{th}\, \forall k \label{constraint:user.MC.SNR}}
    \addConstraint{\hspace{-0.2in}q[0] = q_0 \label{constraint:user.MC.q0}}
    \addConstraint{\hspace{-0.2in}q[K] = q_f, \label{constraint:user.MC.qf}}
\end{mini!}
where, $\Delta t$ is the length of each time step, $\mathcal{E}_{\mathcal{K}}(\{q[k]\})$ is the motion energy required to follow trajectory $\{q[k]\}$, as determined by the UV's energy model (see Section~\ref{sec:modeling}), $\mathcal{E}_{\text{com}}$ is the communication energy, $\Delta t K$ is the mission completion time, and $\lambda_i,\, i\in\{1,2,3\}$ are scalar weights. Thus, the objective consists of a weighted sum of motion energy, communication energy, and mission completion time, with the latter particularly important in, \textit{e.g.}, drone delivery problems. We can formulate many related problems by moving performance metrics between constraints and objectives. For example, the SNR constraint can be relaxed by including a penalty term in the objective for the distance or time traveled with $\text{SNR}<\text{SNR}_{\text{th}}$, and the mission completion time can be recast as a constraint on the maximum allowed time.

A version of this problem is studied in \cite{ICC2021Hurst}, in which a UGV moves through a complex environment with obstacles while maintaining connectivity to the core network via any one of several TBSs. Both fixed and variable transmit power problems are considered, and due to the additive nature of the objective function, the problem can be solved using a modified version of $\text{RRT}^*$. The resulting trajectories conserve communication energy by avoiding areas of poor channel quality. In a similar setting, \cite{TC2019Zhang} studies the problem of a UAV flying between an initial and final location in minimal time while maintaining connectivity with a set of TBSs. The authors find high-quality approximate solutions using tools from graph theory and convex optimization.

When the UV may choose from among multiple BS for connectivity, it should avoid excessive handoffs, which are expensive from the standpoint of communication resources. This is accounted for in \cite{ICC2021Cherif}, which studies the problem of a cargo UAV flying at a fixed altitude between two points which seeks to minimize a linear combination of energy consumption and number of handoffs while constrained to be disconnected no more than a given percentage of the path. The path planning problem is solved using a dynamic programming approach. As the flight altitude increases, the energy consumption increases and handoffs decrease, while stricter disconnectivity constraints also increase energy consumption.

\textbf{Takeaways:}
Maintaining connectivity can be viewed as a shortest-path problem with an appropriate choice of distance metric. Consequently, graph theory and DP provide good approaches to solving these problems. The resulting paths improve communication by seeking out areas of high channel quality, possibly at the expense of additional motion energy. 

Overall, these techniques will allow for continuous HRLLC, a core featyre of 6G systems, which will be needed to maintain command and control of UVs, even in complex environments. Furthermore, as communication channels become quasi-optical, planning paths that maintain connectivity can become more geometrical in nature. Finally, UVs may also need to rely on RIS due to the high-penetration loss at mmWave and higher frequencies.

\subsubsection{\textbf{Minimizing Distance to Connectivity with Stochastic Channel Models}}
\begin{figure}
    \centering
    \includegraphics[width=3.4in]{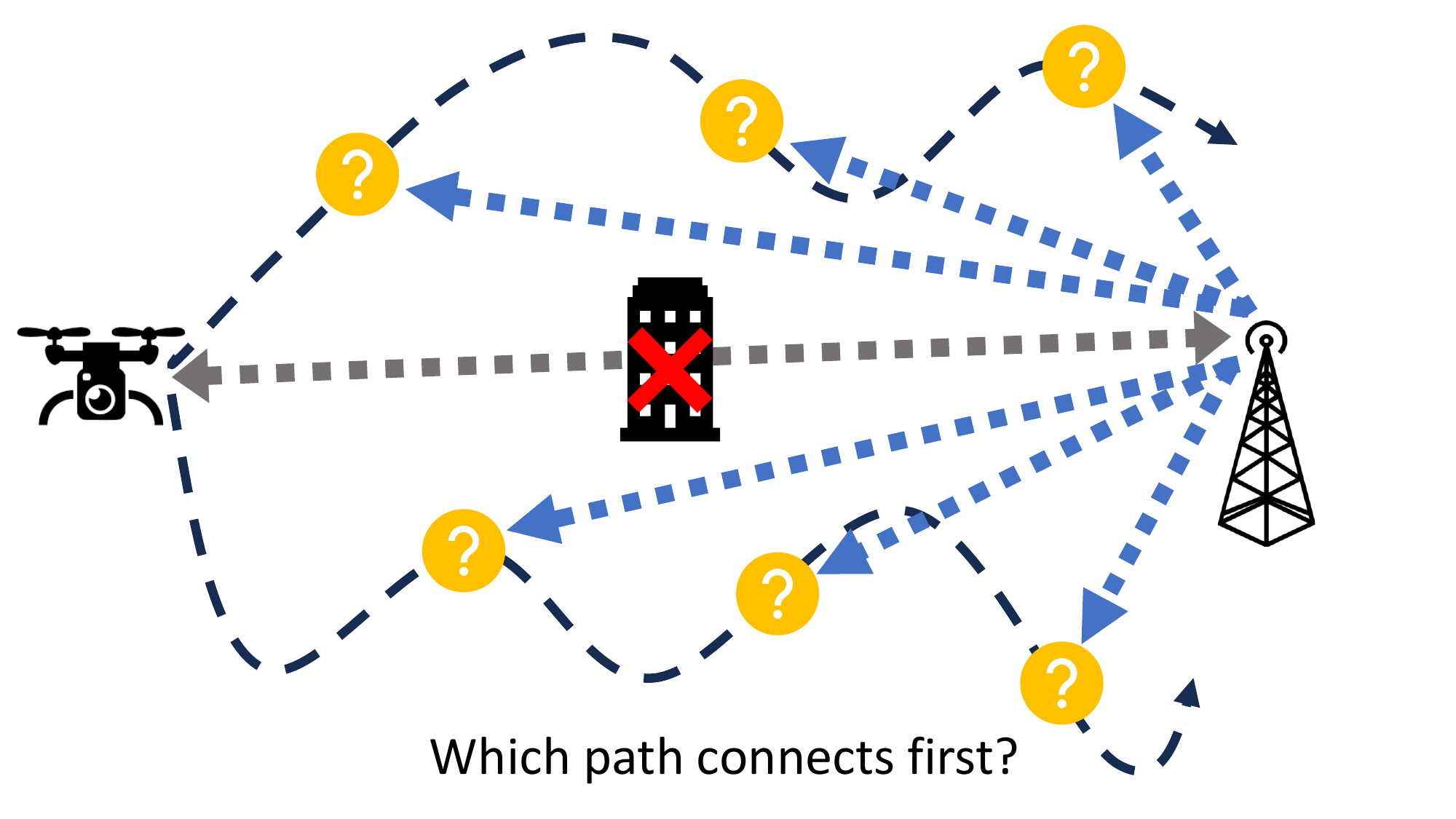}
    \caption{Finding a connected spot: For a UV that is initially disconnected, planning a trajectory that minimizes the time to restore connectivity is challenging, particularly when the channel is not fully known.}
    \label{fig:shortest_to_connectivity}
    \vspace{-0.1in}
\end{figure}

When a UV loses connectivity, either while fulfilling a mission objective or due to the difficult-to-predict variations in the communication channel, it may need to quickly return to an area where communication is possible. Here, a problem naturally arises: how can the UV minimize the time it takes to find a connected point?

Consider a UV initially at a position $q_0$ where the channel quality does not permit communication with a distant BS. Without loss of generality, assume the UV moves at a constant velocity, $v$, and transmits with a fixed transmit power, $P$. The channel is not known, and at each point, the condition (\ref{constraint:user.MC.SNR}) is met with some probability tied to the spatially-correlated statistics of the underlying channel. Further suppose there exists a known, non-empty set of points, $Q_\text{f}$, where, with probability 1, the channel is sufficiently strong to allow for communication when the UV is at any of those points. The problem of minimizing the distance traveled to connectivity can be posed as follows:
\vspace{-0.1in}
\begin{mini!}|s|[2]                   
    {\substack{\{q[k]\}_{k=1}^K, K}}                
    {\mathbb{E}\bigg[\sum_{k=0}^{k_\text{fpd}-1}  \|q[k] - q[k+1]\|_2 \bigg] }   
    {\label{problem:user.FPD}}             
    {}                                
    \addConstraint{\text{Constraints }(\ref{constraint:user.MC.q0}), (\ref{constraint:generic_kinematics})}\notag
    \addConstraint{q[K] \in Q_f, \label{constraint:user.MC.Qf}}
\end{mini!}
where $k_\text{fpd}$ is defined as $\min \{k\,|\,\text{(\ref{constraint:user.MC.SNR}) holds}\}$, and, as $k_\text{fpd}$ is a random variable due to imperfect channel information, the expectation is taken with respect to $k_\text{fpd}$. Importantly, Constraint (\ref{constraint:user.MC.Qf}) ensures the expectation is always well-defined.

The problem of minimizing the expected distance to connectivity in a general sense is studied in \cite{TR2019Muralidharan}. A graph-theoretical formulation is put forth, where at each node, the UV has some probability of success, and the UV must plan a path through the graph which minimizes the expected distance traveled until success occurs. A Markov decision process describes the problem, which is proven to be NP-hard. To find high-quality approximate solutions, the problem is recast as a potential game with the nodes acting as the players, and log-linear learning is used to find near-optimal solutions. Results show that first seeking out nearby locations with even modest probabilities of connectivity can reduce average distance traveled compared to moving directly to the nearest connected point.

When the UV's trajectory is given beforehand, \cite{Autonomous19_MuralidharanMostofi} provides rigorous analysis of the first-passage distance (FPD), the expected distance the UV will travel before reconnecting. An Ornstein-Uhlenbeck \cite{JAP1988Ricciardi} process is shown to model log-normal channel shadowing along a straight-line path, which permits an analytical expression for the FPD. If small-scale fading is not negligible, a recursive formulation is developed that allows the FPD to be calculated over a discretization of the path.

\textbf{Takeaways:}
These problems have a similar shortest-path element as those in (\ref{problem:user.MC}), but when accounting for shadowing and path loss statistically, they become \textit{stochastic} shortest path problems \cite{DPandOC2017Bertsekas}, which are much more difficult to solve. Overall, solving these problems enable ubiquitous connectivity, which is crucial for realizing the 6G vision.

\subsubsection{\textbf{Scheduling Wireless Data Uploading}}\label{sec:PF.consumer.wdu}

\begin{figure}
    \centering
    \includegraphics[width=3.4in, trim={0in, 1.25in, 0in, 0.6in}, clip]{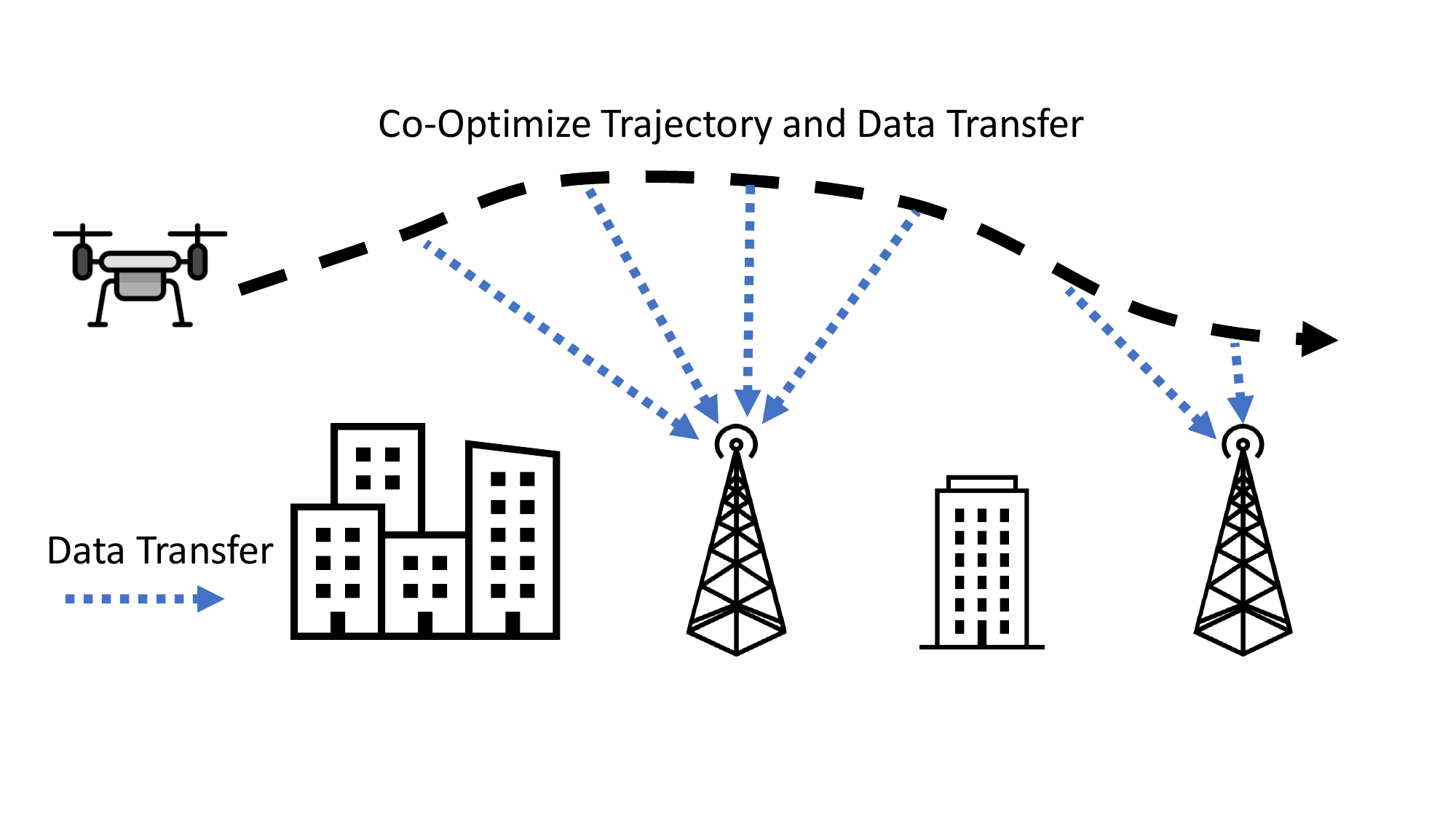}
    \caption{Scheduling of wireless data offloading from a UV to a BS with complementary trajectory design can be a challenging problem, but it can also significantly reduce the energy needed for data transfer.}
    \label{fig:data_offloading}
\end{figure}

In a number of scenarios, UVs must offload data to a remote station while traveling. In these situations, the UV should jointly optimize its trajectory and the corresponding adaptive transmission power/rate over the whole trajectory, for efficient data transfer. The general problem scenario is shown in Fig.~\ref{fig:data_offloading}.

A core problem involves a UV moving from $q_o$ to $q_f$, while offloading $D_0$ bits of data along the way. The problem can be formulated starting with Problem (\ref{problem:user.MC}) and replacing the SNR constraint (\ref{constraint:user.MC.SNR}) with a data offloading constraint:
\begin{subequations}\label{constraint:user.DO.data_offloading}
\begin{alignat}{2}
  &D[0] = D_0,\label{constraint:user.DO.data_offloading.a}\\
  &D[k] = \text{max}(D[k-1] - \Delta t\,B\,r(P[k],\,q[k]),0),\label{constraint:user.DO.data_offloading.b}  \\
  &D[K] = 0, \label{constraint:user.DO.data_offloading.c}
\end{alignat}    
\end{subequations}
where $B$ is the bandwidth of the communication channel, $r(P,\,q)$ gives the data rate between the UV and the BS when the UV transmits with power $P$ from position $q$. Alternatively, an additional term capturing the timeliness of data offloading can be added to the objective.

Scheduling when to upload data adds complexity, particularly for optimization of the transmit power. In problem (\ref{problem:user.MC}), it is easy to see that transmit power should be chosen to satisfy Constraint (\ref{constraint:user.MC.SNR}) with equality, so that the choices of transmit power, $P[k]$, are not directly coupled over time. However, the introduction of Constraint (\ref{constraint:user.DO.data_offloading}) results in tighter coupling of transmit power across time. We next present interesting studies related to this problem.

For a UGV moving along a predefined trajectory, \cite{TWC2013Yan} considers jointly planning data transmission (\textit{i.e.}, communication rate/power) and velocity to minimize total energy costs. It is shown that the UV should move quickly through areas of poor communication channel quality and more slowly through regions of high channel quality (via an optimized velocity profile), so that it spends more time transmitting in more favorable areas. Using the channel prediction framework of \cite{TWC2011Malmirchegini}, an adaptive, online strategy is proposed. The proposed method significantly reduces energy consumption compared to the baseline where the UV moves with constant speed and transmits with constant spectral efficiency.

A continuous time, energy-minimizing version of the problem appears in \cite{ACC2015Ali}, where it is reformulated by moving Constraints (\ref{constraint:user.MC.qf}) and (\ref{constraint:user.DO.data_offloading.c}) into the objective as additional penalty terms. This puts it in the form of an abstract Bolza optimal control problem \cite{Bolza2013Beckenbach}. This work is extended to a multi-UV case in \cite{TCNS2018AliCai}, which proposes a distributed version of the optimal control solution in a multi-agent scenario. Results show that the UVs seek out areas of better channel quality and perform most of the wireless data uploading there.

\textbf{Takeaways:}
A key feature of 6G systems is the intelligent allocation of resources with the end goal of increased sustainability. Jointly adapting communication rate and UV speed to channel conditions can greatly improve communication energy efficiency. This may motivate planning trajectories that seek out areas of better channel quality, though the associated motion energy expenditure must be considered as well.

Overall, the ubiquitous, highly reliable and low latency networks envisioned in 6G will better handle larger amounts of data. Thus, several emerging paradigms can be addressed by these problem formulations. In addition, if the collected data is used in edge computing, some of the constraints will become related to edge-based resources, in addition to bandwidth-related resources.

\subsubsection{\textbf{Site-Visiting Problems}}

\begin{figure}
    \centering
    \includegraphics[width=3.4in, trim={0in, 0.6in, 1.2in, 0.4in}, clip]{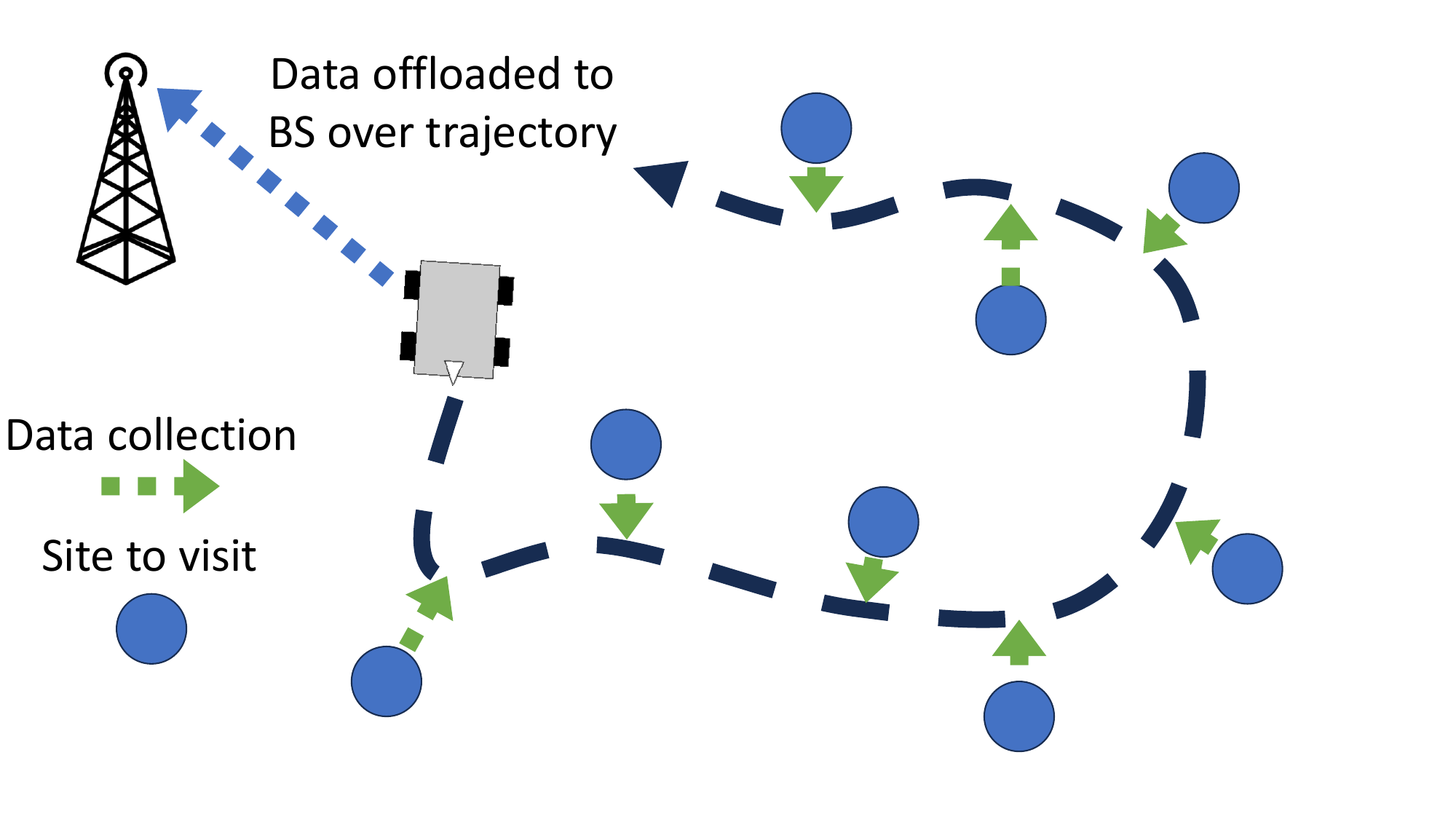}
    \caption{Site visiting problems arise in a number of situations, including surveillance, surveying, and IoT/WSN systems. In addition to collecting data from the sites, the UV may also need to transfer the data back to a remote BS in a timely manner.}
    \label{fig:site_visiting}
\end{figure}

As a final set of problems, we consider scenarios where the UV visits several sites to collect data either with onboard sensors, such as cameras, or through communication channels, e.g., in the case of data collection from wireless sensor nodes or IoT devices. Additionally, the problem may involve transferring the collected data to a remote BS, either over a wireless communication link, in which case the UV's role resembles that of a relay, or by physically moving the data to an access point, sometimes referred to as data muling \cite{SNPA2003Shah}. A general problem scenario is illustrated in Fig.~\ref{fig:site_visiting}.

Usually, site visiting constraints ensure that all sites are visited, and the visit order must be determined, which introduces a combinatorial aspect to these problems. Site visiting constraints can take many forms. At the simplest, they constrain the trajectory to pass through a (possibly singleton) set $\mathcal{Q}_\omega$, e.g., there exists a point in the trajectory, $q[k]$, where $q[k]\in \mathcal{Q}_\omega$ for all $\omega$ in the set of sites $\Omega$. When sites correspond to the locations of IoT devices or sensor nodes, the visiting constraint may ensure that a set amount of data is uploaded to the UV.

Site visiting problems also bring with them a distinct set of metrics, which may enter the system either as constraints or objectives. When the sites correspond to communication nodes, the energy required to upload the data to the UV is an important measure of system performance, as these nodes are frequently quite energy-limited. Additionally, metrics such as average delay and Age of Information (AoI) \cite{TxIT2017Sun} capture the timeliness of data transfer to the BS.

A general site visiting problem can be expressed as:

\begin{mini}|s|[2]                   
    {\substack{\{q[k]\}, \mathcal{R}_{UV},\\\mathcal{R}_\Omega}}                
    {f_\text{GLB}\big(\;\mathcal{E}_\text{UV},\, \{\mathcal{E}_\omega\}_{\omega\in \Omega},\,f_T(\{q[k]\}, \mathcal{R}) \big) }   
    {\label{problem:user.QoI}}             
    {}                                
    \addConstraint{\hspace{-0.3in}(\ref{constraint:generic_kinematics})
    \;\text{Motion constraints}}
    \addConstraint{\hspace{-0.3in} g_c(\{q[k]\}, \mathcal{R}) \leq 0}\;\text{Comm. res. constraints}
     \addConstraint{\hspace{-0.5in}  \begin{rcases}
    &g_{D}(\mathcal{R}, \{q[k]\}) \leq 0\\
    &h_{D}(\mathcal{R}, \{q[k]\}) = 0
    \end{rcases}\,\begin{array}{l} \text{Site Visiting}\\ \text{Constraints}\end{array}},
\end{mini}
where $\mathcal{E}_\text{UV}$ is the total energy consumed by the UV, $\mathcal{E}_{\omega}$ is the energy expended by site $\omega$ to upload data to the UV, $\mathcal{R}_{UV}$ and $\mathcal{R}_\Omega$ are the communication resource variables for the UV and sites, respectively, with $\mathcal{R} = \mathcal{R}_{UV} \cup \mathcal{R}_\Omega$, and $f_T(\{q[k]\}, \mathcal{R}_{UV}, \mathcal{R}_\Omega)$ is a measure of the timeliness of data collection, such as mission completion time or AoI.

\textbf{Data Collection for IoT/WSN:} A number of works focus on the problem of data collection from IoT/WSN devices without explicitly modeling the data transfer from UV to BS. In these cases, the energy consumed by the devices, $\mathcal{E}_\omega$ is often given greater attention. For example, \cite{TWC2017Mozaffari_IOT} studies the problem of deploying multiple UAVs to collect data from IoT devices in a way that minimizes energy expenditure by the devices. Using BCD and sequential quadratic programming, the authors find device-to-UAV associations, UAV position, and device transmit power schedules that greatly reduces the total energy consumed by the IoT devices compared to the scenario where the UAVs are statically placed in the environment.

In \cite{TWC2020Samir}, the authors consider the challenging problem of multiple UAVs collecting data from IoT devices under time constraints, that is, when the devices are only awake for a subset of the time. They formulate a mixed-integer nonlinear program (MINLP) to find a trajectory and an allocation of communication resources which maximize the total number of IoT nodes serviced over a finite time frame. A similar time-constrained, multi-UAV problem is considered in \cite{GLOBECOM2020Ghdiri}, which instead focuses on the problem of minimizing the total UAV energy consumption while ensuring that a set amount of data is collected from each node. The problem is solved using results from the literature on capacitated vehicle routing problems with time windows (CVRPTW).

\textbf{Sensing and Data Transfer:} A different set of problems focus on the UV-to-BS offloading aspect of these problems. These problems are similar to the data offloading problem in Section~\ref{sec:PF.consumer.wdu}, but with modified data dynamics. Specifically, after visiting each site, an additional amount of data enters the UVs onboard memory, so that Eq. (\ref{constraint:user.DO.data_offloading.b}) is replaced by
\begin{equation*}
    D[k] = \text{max}\big(D[k-1] + \sum_{\omega\in\Omega} I_{\omega}[k] D_\omega - \Delta t\, B \,r(P[k], q[k]),\,0\big),
\end{equation*}
where $I_\omega[k]$ indicates if site $\omega$ is visited at time step $k$ and $D_\omega$ is the data associated with site $\omega$.

In \cite{TCNS2014Yan}, a UGV visits a number of sites where it collects data and transmits it to a remote station. Initially, the problem is considered in the absence of the sites, and the UV simply decides whether it is more energy efficient to transmit from the current location or to first expend energy to move to another location and then transmit. Then, the problem of minimizing energy consumption by co-optimizing the UV's motion and communication strategies for retrieving and relaying data from the site back to a BS is formulated as a MILP. 

A related problem is studied in \cite{ACMTSN2014Ghaffarkhah}, in which multiple UGVs service a number of sites to keep a generic Quantity of Interest (QoI) from growing unbounded while maintaining communication with a remote BS. Two cases of connectivity are considered. In the first, the UVs must establish a communication link to a remote BS after each site visit. In the second, the UVs must only connect once during their cyclic trajectories. The problem of minimizing the total energy consumption while guaranteeing boundedness of the quantity as well as respecting the connectivity constraints is formulated as a MILP.

\textbf{Data Collection and Relay:} A final set of problems considers uplink communication both from the sites to the UV and from the UV to the BS. These types of problems are the most general and frequently feature metrics related to the timeliness of end-to-end data transfer. For example, \cite{TWC2023Hurst} studies the problem of minimizing the average wait time in a data collection and relay scenario. The authors consider a single UGV relaying data between a set of spatially diverse and disparate source-destination pairs, where heterogeneous data accumulation rates at the sources necessitate different visiting frequencies. They then find the optimal robotic routing policies (e.g., optimal relay locations as well as the sequence with which the pairs are serviced). More specifically, the problem of minimizing the average time a bit of data waits at the source before being relayed to the destination is approached from a polling systems perspective \cite{Mathematics2021Vishnevsky}, and several fundamental properties of the system are proved using results from queuing theory, all the while showing how stochastic robotic routing can be handled.

Another metric of the timeliness of data transfer is Age of Information (AoI) \cite{TxIT2017Sun}, which measures the recency of the most up-to-date data at the BS. In \cite{TWC2023Gao}, the authors look at the problem of minimizing the AoI in a scenario where multiple UAVs are dispatched to collect data from clusters of devices and then return to a data center to offload the data. The AoI accounts for the travel time as well as the device-to-UV and UV-to-BS communication times. The problem is approached by first partitioning the sites among the UAVs with a variant of K-means \cite{Electronics2020Ahmed}, then using ant colony optimization \cite{ACO2005Blum} to find the visit order for each UAV. For a single UAV, end-to-end AoI is also addressed in \cite{TVT2019Abd-Elmagid}, but the data transfer to the BS is constrained to happen directly after receiving the data from the device, that is, the UAV operates as a half-duplex relay. The authors use BCD along with SCA and convex optimization methods to produce trajectories and transmit power profiles which minimize the average AoI while respecting communication energy constraints for the UAV and devices.

\textbf{Takeaways:}
The combinatorial element of site visiting and wireless offloading with causality constraints together produce very complex problems with elements of both continuous and discrete optimization coupled strongly across time. Furthermore, when the UVs collect data via wireless uploading from devices, additional considerations come into play, including the energy consumed by the devices and possibly timing windows. For end-to-end data transfer with data generating dynamically at each site, the system may be viewed as a network of queues \cite{SIAM1985Disney}, where data accumulating in queues at the sensor nodes are uploaded to the queue on the UV which in turn uploads the data to a BS. The system also bears resemblance to a polling system, in which a single server serves multiple queues, with some idle time incurred when switching between the queues.

Overall, these types of site-visiting problems can play a key role in the 6G eco-system. They can even be handled by teams of possibly heterogeneous UVs~\cite{Zhu2023APSCON}, which will add another interesting layer of complexity to the problem formulation.

\section{Stochastic Dynamic Programming Formulations}\label{sec:rl_formulations}

\textcolor{black}{Many of the analytical formulations of the previous section (along with the accompanying solution methods) account both for the inherent stochasticity of UV-assisted wireless systems and for the optimization of system parameters over an extended period of system operation. Nonetheless, the recent paradigm of DRL introduces a compelling framework for advancing these formulations. In a dedicated subsection, we delve into the evolving landscape of stochastic dynamic programming formulations within UV-assisted networking. Furthermore, we explore the application of learning-based methods (and more prominently DRL) to effectively address such settings. }

\subsection{Unmanned Vehicles as Providers}

The current subsection provides an overview of UV problem formulations that adhere to the general framework of stochastic dynamic programming and consider the UV as a provider of system services. In such a role, the UV can be a central node that provides connectivity with the core network, e.g., an aerial base station (BS). Alternatively, the UV can act as a relay node that extends coverage and connectivity.

\subsubsection{UVs as Base Stations}
Assume a UV acts as an aerial BS and serves $\mathcal{U}$ ground users located at specific positions. The elevation level of the UV stays constant throughout the operation time horizon. The UV starts at position $q_{0}$ and must finish its flight trajectory at position $q_{K}$ at time $K$. The channel at time $k$ between the UV and the ground user $i$, $\mathcal{H}_{i}$,
depends on the position of the UV $q[k]$, $\mathcal{H}_{i}(q[k])$. The transmission power of the UV at time step $k$ is denoted as $P[k]$. The communication rate for user $i$ at time step $k$ is $\log_{2} ( 1 +  \text{SNR}(\mathcal{H}_{i}(q[k]))$. The set of all valid trajectories in the UV operation space that amount to overall time $T$ is denoted as $\mathcal{Q}$. The problem of maximizing the expected sum-rate is formulated as follows:

\begin{maxi!}|s|[2]
     {\mathcal{Q}}                              
    {\mathbb{E}_{\mathcal{Q}} \left[ \sum_{k=0}^{K} \sum_{i \in U}  \log_{2}\left( 1 + \text{SNR}(\mathcal{H}_{i}(q[k]) \right) \right]}   
    {\label{opt:UAV_relay} }             
    {}                                
    \addConstraint{q[0]=q_{0}}
    \addConstraint{q[K]=q_{K}}
    \addConstraint{q[k] \in [q_{min}, q_{max}]^{2} \hspace{0.3cm} \forall k \in [0, K]}
    \addConstraint{\log_{2}(1 + \text{SNR}(\mathcal{H}_{i}(q[k])) \geq \zeta_{min} \hspace{0.3cm} }
    \addConstraint{\sum_{k=0}^{K} P[k] \leq P_{total}.}
\end{maxi!}

The goal is to design the UV trajectory  over time such that the sum of user communication rates is maximized in expectation. The constraints correspond to the fact that the initial and final positions of the UV are predetermined, the UV area of motion is confined within a $2D$ rectangular space, and there should be fairness among end users in the sense that the communication rate of any individual user is not below a certain threshold at all time steps. Finally, the overall power budget for transmission is below a certain predefined threshold.

The problem formulation calls for approaches where optimization unfolds through time. DRL is an organically suitable paradigm for tackling the problem. To employ DRL for settings that fall under the above formulation, the Markov decision process (MDP, see Section~\ref{sec:approaches.DRL}) must be defined:
\begin{itemize}
    \item \underline{State}: $\mathbf{s}[k] = f_{s}(q[k])$, the state is a vector that is a function of the UV position at time step $k$. In the simplest case, the state can be the position itself, or a window of previous positions. 

    \item \underline{Action}: $\mathbf{a}[t] = f_{a}(\delta q[k])$. The action is a function of the displacement of the UV for time step $k$. The displacement can be a vector that represents the difference between the position at the current step and the position in the subsequent step. The range of the displacement should adhere to the imposed kinematic constraints.

    \item \underline{Reward}: $r[k] =  \sum_{i \in U}  \log_{2}\left( 1 + \text{SNR}(\mathcal{H}_{i}(q[k + 1]) \right)$. The reward is the overall communication rate achieved at time step $k+1$. One should note that, by the definition of the MDP, the reward is calculated after performing the action at the current state. To facilitate the fairness constraint, the reward can be augmented by the difference between the minimum communication rate among end users at each time step and the minimum acceptable communication rate $\zeta_{min}$ and the transmission power at every time step.
\end{itemize}

A significant body of work exists that falls under the aforementioned formulation. The work in \cite{bayerlein-IEEESPAWC-2018trajectory} examines a scenario where a UV acts as a BS (UV-BS), and the goal is to optimize the total communication rate over time. A Q-learning \cite{watkins-SpringerML-1992q} approach is considered, and the space of UV motion is discretized into a fine grid. In \cite{qin-TVT-2021distributed}, the authors focus on a similar scenario with multiple UV-BSs, where the goal is to maximize a weighted throughput objective that relates to fairness among end users. A Multi-agent Deep Deterministic Policy Gradient approach is proposed that disentangles the contributions of the individual UVs. The authors in \cite{mohammadi-Arxiv-2022analysis} examine a setting where a UV-BS serves end users in the context of Open Radio Access Networks (ORAN) \cite{singh-IEEEWCNC-2020evolution} under MIMO fading channels. The UV motion is discretized, and two RL approaches are proposed to maximize the network's throughput over time. The first approach is a variant of SARSA \cite{zhao-IEEESSCI-2016deep}, and the second approach is predicated on offline Q-learning \cite{kostrikov2021offline}.

The work of \cite{zhan-IEEETWC-2022energy} examines the trajectory optimization of multiple UVs that serve a ground cellular network. The goal is to design the trajectories of the UV swarm and the task completion time in order to minimize energy expenditure, subject to a constraint that pertains to the continuous connectivity of the UVs with the ground cellular network. The authors consider two scenarios: one where the radio channel maps are known to the swarm and the other where the channel maps are initially unknown. For the scenario where the channel maps are known, they employ a convex optimization approach along with a dynamic-weight shortest path algorithm. For the second scenario, the aforementioned approach is non-applicable, and the authors propose a deep Q-learning approach with a dueling architecture in order to optimize the UV trajectories over time. The work of \cite{zeng-IEEEGLOBECOM-2019path} examines a multi-UV cellular network, where the UV trajectories are optimized with a TD-learning \cite{tesauro-ACMCOM-1995temporal} approach to minimize the overall task completion time. The state-action value function is linear and tile-coding \cite{sherstov-SpringerISARA-2005function} is implemented to ameliorate the effect of the large state-action space. A similar scenario is considered in \cite{yin-IEEETVT-2019intelligent}, with the difference being that only one UV is considered. A Deterministic Policy Gradient algorithm is developed to optimize the UV motion with the goal of maximizing the uplink sum-rate.

\subsubsection{Cooperative Communication} Besides the UV operating as a BS, there has been increasing interest in applications where the UVs are deployed as intermediate nodes in the cascade of the network \cite{zeng2016throughput}. The inherent mobility of UVs can be leveraged to increase coverage and improve the QoS on the receiver side.

Consider $I$ UVs acting as relays that facilitate the communication between a source $S$ and a destination $D$. The locations of source and destination are fixed throughout the operation, and therefore the dependence on their location is dropped from all expressions. The source transmits symbol $s[k] \in \mathbb{C}$ with power $P$, and the signal
received at UV-relay $i$, located at position $q_{i}[k]$,  depends on the position of the relay and the channel from the source to the UV-relay. The channel from the source to the relay, $\mathcal{H}_{Si}$ also depends on the relay position and possibly on the time (assuming spatio-temporally correlated channels \cite{kalogerias2017spatially}). The signal received by relay $i$, $x_{i}[k]$ can be generally expressed as:
\begin{equation}
    x_{i}[k] = x_{i}\left(q_{i}[k], \mathcal{H}_{Si}[k]\right) .
\end{equation}
When considering narrowband flat fading channels, the effect of the channel on the transmitted signal is multiplicative \cite{TSP2022Evmorfos, kalogerias2017spatially}. On the other hand, wideband channels of 6G systems are frequency selective \cite{wang2022towards,tan2019delay} (see Section \ref{sec:modeling.channel}).

Each of the UV-relays multiplies its received signal with weight $w_{i}[k]$, and all UVs collectively transmit their signals to the destination using a collective power budget $P_{I}$. The channel from each UV to the destination, $\mathcal{H}_{iD}$, depends on the UV's position and the time instance. The received signal at the destination $x_{D}$ is a function of all channels (from the source to the UV and from the UV to the destination) and all relay positions:
\begin{equation}
    x_{D}[k] = \mathcal{F}\left(\{ q_{i}[k]\}_{i=1}^{I}, \{\mathcal{H}_{Si} \}_{i=1}^{I}, \{\mathcal{H}_{iD} \}_{i=1}^{I} ,k\right),
\end{equation}
where $\{\mathcal{H}_{Si} \}_{i=1}^{I}$ denotes the channels from the source to all deployed UVs/relays and $\{\mathcal{H}_{iD} \}_{i=1}^{I}$ the channels from all relays to the destination.  Noise is induced at both the signal reception at each UV and the signal reception at the destination. Therefore, the destination signal can be decomposed as:
\begin{equation}
    x_{D}[k] = x_{D}^{signal}[k] + x_{D}^{noise}[k],
\end{equation}

\textcolor{black}{Under the assumption that the channels (both from the source to each UV and from each UV to the destination) exhibit correlations with respect to time and space, the problem can be formulated as:}
\begin{maxi!}|s|[2]
     {\mathcal{Q}}                              
    {\mathbb{E}_{\mathcal{Q}}\left[\sum_{k=0}^{\infty}\mathcal{U} \left( \{\mathcal{H}_{Si}[k] \}_{i=1}^{I}, \{\mathcal{H}_{iD}[k] \}_{i=1}^{I}\right)\right]}   
    {\label{opt:UAV_relay_dp} }             
    {}                                
    \addConstraint{q_{i}[0]=q_{i0} \hspace{0.2cm} \forall i}
    \addConstraint{||q_{i}[k+1] - q_{i}[k]||_2\leq v_\text{max} \Delta_k \hspace{0.2cm} \forall i,k}
    \addConstraint{\mathbb{E}\left[||x_{i} [q_{i}[k], k]||^{2}\right] \leq P_{I}.}
\vspace{-0.2cm}    
\end{maxi!}

The set of all infinite duration permissible trajectories of UVs is denoted as $\mathcal{Q}$. $\mathcal{U}$ is a general utility function of the system that depends on the realization of the channels and therefore on the corresponding relay positions for all time steps of operation. The most typical choice for the utility is the SINR at the destination, which is expressed as:
\begin{equation}
   \mathcal{U}[k] \equiv \text{SINR}[k] = \frac{|x_{D}^{signal}[k]|^2}{|x_{D}^{noise}[k]|^2}.
\end{equation}
Besides the SINR, there exists a plethora of other candidate utility metrics, such as the sum-rate (for multiple receivers) and the secrecy rate (refer to Section \ref{subsec:utility_metr}). The choice of the utility function introduces different nuances that are domain-specific and have to be explicitly addressed (e.g., additional interference when considering multiple destinations). 
The above formulation paves the way for DRL approaches to be adopted. The works \cite{TSP2022Evmorfos, Evmorfos-IEEEAsilomar-2021doubledeep, Evmorfos-IEEESPAWC-2021deepq} consider multiple UVs as relays facilitating the communication of a single source-destination pair. The space over which the UVs can move is discretized into 2D grid cells, and various variations of the deep Q-learning approach \cite{mnih-Nature-2015human, wang-ICML-2016dueling} are proposed to solve the problem. In those works, the authors note the inability of Multilayer Perceptrons (MLPs) with ReLU activations to capture the high frequency spectral components of the optimal value function, which arise due to the pronounced variability of the underlying channels with respect to time and space. It was also noted that by preprocessing the input of the state of the corresponding MDP with a learnable Fourier kernel, those issues can be mitigated, leading to an increase in convergence speed and average SNR accumulation at the receiver.
In \cite{ICASSP2022Evmorfos}, the assumption for space discretization is relaxed, and a variation of the Soft Actor-Critic \cite{haarnoja-ICML-2018SAC} algorithm for continuous motion control is proposed.

The works \cite{Kalogerias-IEEETSP-2018Spatially, Kalogerias-IEEEICASSP-2016mobile} also examine the aforementioned problem and propose a stochastic programming approach to solve it. The approach is data-driven and optimizes the SNR in expectation for the subsequent time step of system operation. In \cite{Evmorfos-FrontiersSP-2022adaptive}, the authors perform a comparative analysis of the performance of the DRL approach of \cite{TSP2022Evmorfos} and the stochastic programming approach of \cite{Kalogerias-IEEETSP-2018Spatially}. Finally, the work \cite{Diamantaras-MLSP-2019Optimal} considers the same problem and develops a multi-armed bandit method to tackle it.

{The work \cite{shin-IEEEWCL-2023sub} studies a scenario where a single UV serves multiple users as a relay. Multiple model-free RL algorithms are proposed for UV motion control, with the goal of maximizing energy efficiency under quality-of-service constraints. The proposed RL approaches are shown to surpass the performance of a connected hybrid precoding approach. The work of \cite{abohashish-SpringerWCN-2023trajectory} examines scenarios where a UAV serves as a mobile relay that maintains the communication between multiple destinations (end users) and a BS. Both model-free and model-based RL approaches are proposed for the same setup. In \cite{hassan-IEEEICC-20223to}, the authors examine a scenario where one UAV acts as a relay and  propose an on-policy DRL approach for trajectory design, with the goal being to optimize the average throughput of the collective system. The DRL approach is a proximal policy optimization method \cite{schulman-Arxiv-2017proximal}. The proposed algorithm accounts for the molecular absorption effect that is prominent in THz communications \cite{han-IEEEWC-2022molecular}.}

The work of \cite{wang-IEEECL-2019multi} examines a communication set up where multiple UVs serve as relays to facilitate the communication between a BS and multiple end users. A deep Q-learning approach with a dueling architecture of the Q-network \cite{wang-ICML-2016dueling} is proposed to optimize the UV locations over time. The goal is to design the trajectories of all deployed UVs in order to maximize the downlink capacity of the network under the constraint that all end users should be served by at least one relay at all times of system operation. 
The work by \cite{9766100} addresses the joint path planning optimization of a swarm of UVs. The authors propose a multi-agent DRL approach that incorporates an attention mechanism for function approximation. They conduct an empirical analysis comparing the effects of multi-head versus single-head attention in scenarios with partial observability induced by agent cooperation. Their findings suggest that using single-head attention can reduce completion time and improve overall communicatio efficiency. The authors in \cite{9415745} also consider a multi-UV setting for Machine-to-Machine (M2M) communications. They propose a DRL approach that is assisted by a Generative Adversarial Network (GAN) with Long-Short-Term-Memory (LSTM) units in order to predict future interactions. The employment of the GAN provides considerable improvement in sumrate.

\textbf{Heterogenous Multi-agent RL:} An important consideration in using UVs for cooperative communication is the need to account for the heterogeneity in UV characteristics when they serve as intermediate nodes for wireless connectivity. Heterogeneous UVs vary in constraints and capabilities, which is crucial for network flexibility. This flexibility allows new UVs to replace older ones or be added to the network, thereby expanding its capacity. However, explicitly modeling heterogeneity is challenging. Therefore, data-driven approaches that incorporate the multi-agent aspect are promising for effectively managing this heterogeneity. The study by \cite{9980391} focuses on a network of heterogeneous UVs tasked with completing jobs that arise in real-time. The authors introduce a multi-agent DRL approach with prioritized experience replay, specifically designed to handle the heterogeneity of UVs. This method effectively addresses the challenge of dynamic resource reallocation as new tasks are assigned. The proposed approach outperforms both game-theoretic and single-agent RL methods in terms of efficiency and performance. The study by \cite{10160919} explores a network of heterogeneous UVs tasked with pursuit-evasion operations. The authors introduce a dynamic, role-based multi-agent DRL approach \cite{wang2020rode} that employs Voronoi-based reward functions \cite{hu2020voronoi}. This method significantly outperforms existing baseline approaches in effectively managing the complexities of these tasks.

\textbf{Takeaways:} Both this subsection and the previous one offer insights and problem formulations concerning the intersection between DRL and UV motion control. The distinction between the two lies in the specific roles played by UVs in each scenario. In the first subsection, the UV operates as a provider, akin to a mobile BS, while in the second subsection, it acts as an intermediate mobile node facilitating communication between  source and destination points.

At a cursory glance, the formulations presented in Problem (\ref{opt:UAV_relay}) and the Problem of (\ref{opt:UAV_relay_dp}) might appear similar. Both require the design of UV motion policies to maximize the expected sum of a QoS metric, contingent on the channel conditions, over an extended time horizon. Solving sequential decision-making problems using DRL hinges significantly on estimating the optimal value function of the corresponding MDP. In both cases, this value function depends on the underlying channels. When the UV serves as a provider, only the channel from the UV to the user impacts the value function. Conversely, when the UV acts as a relay or helper, the optimal value function is influenced by both the channels from the UV to the receiver and the channels from the source or BS to the UV. This additional dependency introduces localized variability in the true value function, making its accurate estimation through bootstrapping a challenging endeavor \cite{TSP2022Evmorfos}. These formulations are anticipated to play a pivotal role in 6G, as the next generation of wireless networks is expected to extensively utilize data-driven approaches for dynamic network optimization.

\vspace{0.01in}
\subsection{Unmanned Vehicles as Consumers}
This subsection delves into the collaborative relationship between  DRL and UVs specifically when UVs act as consumers of services within wireless networks. DRL is commonly used for UV  motion control or resource allocation, especially in ``data harvesting" scenarios from IoT/sensor networks. The inherent mobility of UVs is advantageous for efficient data collection from dispersed sensors. However, challenges arise, requiring advanced algorithms to optimize UV motion for objectives like maximizing data throughput, operational time, data freshness, and minimizing energy consumption amidst environmental randomness.

A broad scenario that encompasses various settings of sequential decision-making for  UV-assisted data collection unfolds as follows: A single UV is tasked with collecting data from $M$ clusters of sensors, each cluster comprising the same number of $I$ sensors. The UV's position at each step is represented as $q[k] = \left[x[k], y[k], z[k]\right] \in \mathbb{R}^{3}$. The position of the $i$th sensor within the $m$th cluster at time step $t$ is denoted as $q_{i}^{m} = \left[ x_{i}^{m}[k], y_{i}^{m}[k]\right]$. The throughput from sensor $i$ within cluster $m$ depends on both the UV's position and the sensor's location, under particular assumptions for the underlying channel model: $R_{i,m}[k] = R_{i,m}[q[k], q_{i}^{m}[k]]$.

Considering the existence of a probability of link blockage between the UV and each sensor in every cluster, the sensor's throughput can be defined as $R_{i,m}[k] = P_{i,m}[k] R_{i,m}[q[k], q_{i}^{m}[k]]$, where $P_{i,m}[k] \in [0,1]$ signifies the probability of a lost connection between the UV and the $i$th user within the $m$th cluster at time step $k$. The overarching aim of the sequential decision-making framework is to construct the policy for the UV motion such that the overall throughput achieved throughout the mission is maximized while simultaneously the overall mission time is minimized.

On denoting the set of all permissible UV trajectories by $\mathcal{Q}$, the generic optimization problem can be defined as
\vspace{-0.2cm}
\begin{maxi}|s|[2]
     {\mathcal{Q}}                              
    {\mathbb{E}_{Q} \left[ \sum_{k=0}^{K} \sum_{i}^{I} \sum_{m}^{M} \alpha R_{i, m}[q[k], q_{i}^{m}[k]] - \beta k\right],}   
    {\label{opt:UAV_relay_consum} }             
    {}                                
\vspace{-0.2cm}  
\end{maxi}

\noindent where $\alpha$ and $\beta$ are scalars that manage the design tradeoff between the desire for overall sum-rate maximization and mission completion time minimization.

The work presented in \cite{nguyen-IEEETC-20223d} considers a scenario very similar to the one described above. The state is the position of the UV at each time step, the operational space is discretized into a fine grid, and the action space is comprised by the displacement of the UV to one of the neighboring cells. The authors propose the adoption of the deep Q-learning algorithm to control the UV positions over time in order to optimize a linear combination of the overall sum-rate and the mission completion time. The study presented in \cite{Oubati-IEEETVC-2022Sync} considers a scenario where multiple UVs perform both data collection and wireless energy transfer. In particular, each UV supplies energy to each one of the sensors. Subsequently, the sensors transmit their collected data to the UVs. However, when the same UV is used both for data collection and energy harvesting, both operations are negatively affected. The authors propose splitting the group of UVs into two categories, one that is solely responsible for energy harvesting and another that is solely responsible for data collection. The goal is to optimize the trajectories of both groups of UVs. Since the optimal policies of the two groups are correlated, the authors propose a multi-agent DRL algorithm in order to control the trajectories of the UVs of each category such that a linear combination of the overall throughput, the Age-of-Information, the overall harvested energy and the UV energy expenditure is optimized. The authors of \cite{Li-IEEEIoTJ-2021Joint} examine a challenge, particular to UV-assisted data collection in IoT systems, that pertains to the unsuccessful data transmission due to buffer overflow caused by the maneuvering of UVs. Since the state of the individual IoT nodes can be potentially obscure to the UV, the induced MDP is Partially Observable \cite{spaan2012partially}. The authors propose a deep Q-learning-based approach for Partially Observable MDPs in order to control both the UV flight and the data scheduling with the goal of minimizing the overall packet loss.

A similar data collection problem is considered in \cite{TCNS2022Hong}, which additionally considers the data transfers from the UV to a remote station. The UV's trajectory is co-optimized with an adaptive transmit rate for wireless data transfer to a remote station, and an MDP is formulated, with a cost function which captures both the motion and communication energy of the UV, and a provably-optimal Monte-Carlo tree search (MCTS) method is proposed which treats the optimization of communication resources along the entire path as the final stage in the MDP, thus greatly reducing the size of the state space. In a related work, \cite{CDC2023Hurst} examines the persistent dynamic relay scenario first studied in \cite{TWC2023Hurst},  in which a single UV services a group of disparate source-destination pairs. The authors are then interested in finding the optimal robotic routing locations and dynamic routing policies. The authors show that the problem of minimizing the average wait time can be posed as an average-cost semi-Markov decision process, and they propose using a variation of PPO to efficiently solve the problem, while showing how to handle a long-run average cost criterion.

\textbf{Takeaways:} The successful deployment of UVs in wireless communications as consumers hinges on effectively managing overall heterogeneity. In the context of 6G, these applications are expected to accommodate various types of UVs deployed at different time slots, along with diverse base stations from multiple operators, necessitating inter-operator handovers \cite{CommsMag2020Giordani}. Due to this heterogeneity and network diversity, these challenges are difficult to address through purely analytical methods. Consequently, data-driven adaptive approaches that consider long operational horizons are highly attractive.

\section{Challenges and Opportunities}\label{sec:challenges_and_opp}
In this section, we point out important, high-level challenges and opportunities related to deployment of UVs in 6G systems. The discussion focuses on challenges related to modeling, computation, and security.

\subsection{Towards Comprehensive, High-fidelity Models}
While the models presented in Section~\ref{sec:modeling} will continue to have an important place in optimization and deployment of UVs, emerging aspects of 6G will require advancements and updates in modeling techniques.

\subsubsection{Evolving Models of High-frequency Channels}
Ongoing research into the propagation characteristics of mmWave, sub-THz, and THz frequencies continues to produce new findings, which should be integrated into UV planning~\cite{CST2018Hemadeh, CST2022Han}. This includes appropriate choices of channel parameters based on studies for the specific communication frequency and environments (urban, sub-urban, rural, maritime, etc.) in question. It may also include new ways of thinking about communication channels. For example, as channels become more quasi-optical at higher frequencies \cite{maltsev2009experimental}, channel prediction may become more geometric in nature, and larger channel bandwidths may require new modeling approaches. Furthermore, the development of generative AI for channel simulation may present a possible new approach in some contexts \cite{Chi2024MobiCom_RF-Diffusion}.

\subsubsection{Need for Detailed Kinematics Modeling for 6G Applications}
The advent of 6G and the increased use of high-frequency communication channels can make the simple motion model frequently considered in the literature no longer sufficient. For instance, there may be a need to actively avoid challenges such as airframe occlusion and beam misalignment \cite{zhang2020TVT, wang2024IoTJ}, which require more detailed modeling of both the UV's kinematics and shape. The use of sophisticated simulation offers a potential solution, as discussed next.

\subsubsection{Integrated Models and Digital Twins}
Existing simulation solutions provide either high-fidelity UV kinematics (\textit{e.g.}, Gazebo~\cite{Koenig-2004-394}) or accurate channel modeling (\textit{e.g.}, Remcom Wireless InSite~\cite{remcom}), but not both. The development of such a comprehensive platform that embraces both angles will greatly accelerate research in this area. Along these lines, the use of ``digital twins", simulations informed by real-time data collected from IoT and other devices, offers a possible avenue forward that is well suited for AI and ML techniques \cite{sherazAccess2024}. These would include high-fidelity channel models using, \textit{e.g.}, ray tracing \cite{alkhateeb2023CommsMag} or ML \cite{becvar2023CommsMag}, as well as detailed UV models. However, privacy and security related to sharing the relevant real-time data would require careful consideration.

\subsection{Computational Challenges and Opportunities}
Optimizing UV deployment in 6G faces several computational challenges. While ML techniques offer performance gains, they must be executed with limited computational resources of autonomous vehicles. Larger vehicles may manage this, but smaller UVs may struggle, though advances in chip design \cite{cleary2020AIPR} could help. Furthermore, real-time optimization for large numbers of heterogeneous UVs also poses computational challenges, particularly when they are not part of the same fleet. Centralized optimization is hindered by the scale and latency \cite{mcenroe2022survey}, as well as privacy concerns, while local optimization risks sub-optimal outcomes, especially with non-cooperative or competitive UVs~\cite{tinh2022Access}. On one hand, game theory and mechanism design, used to create decentralized systems with acceptable performance guarantees, offer one approach to these challenges \cite{hurst2024GLOBECOM,tinh2022Access}. On the other hand, edge computing \cite{QU2021FGCS} and federated learning promise new computational paradigms, which we briefly explore next.

\subsubsection{Edge Computing}
UVs can benefit from \cite{mcenroe2022survey} and contribute to edge computing and edge intelligence \cite{liu2022multi}. The effective functioning of UVs, both as consumers and providers of wireless connectivity, relies on the application of computationally intensive algorithms, such as vision algorithms for precise localization and mapping \cite{li2016real}. Leveraging edge servers near UVs allows for substantial storage and computation, striking a balance between latency and computational capabilities. Particularly in the realm of AI on the edge, various research areas such as transfer learning \cite{sufian2021deep}, knowledge distillation \cite{jang2020knowledge}, and neural network pruning/quantization \cite{chandakkar2017strategies} are emerging to facilitate processing on the network edge, in close network proximity to the UVs.

Conversely, UVs can serve as edge nodes where end users offload tasks \cite{zhu2021learning}. The mobility of UVs introduces a dynamic edge computing environment, allowing them to approach users based on a flexible schedule. This dynamic interaction reduces latency and enhances throughput for end users in response to evolving network conditions. Additionally, edge computing brings indirect security benefits to UV operation. As UVs are equipped with onboard sensors such as cameras, they inherently collect data that is private to end users. Leveraging the paradigm of federated learning \cite{cheng2022federated} alongside edge computing helps avoid the transfer and storage of sensitive information on central cloud servers. Instead, storage and computation are offloaded to edge nodes, which are less susceptible to information interception and, therefore, invasion of privacy.

\subsubsection{Federated Learning}
The federated learning paradigm \cite{zhang2021survey} addresses these concerns by keeping data at the terminal device that generated or collected it \cite{qu2021decentralized}. In this approach, a local deep learning model is trained by the data, and only the parameters of the local model are transmitted to the central cloud center. This methodology helps mitigate privacy invasion for end-users. In addition, the transmission of model parameters induces less latency than the transmission of raw data. A plethora of different works leverage federated learning for UV-assisted applications. In \cite{zhang2020federated}, the authors examine distributed image classification for multi-UV exploration. The authors in \cite{wang2020learning}  propose a federated learning scheme that addresses nuanced privacy requirements in UV-assisted crowdsensing. Overall, utilizing federated learning for UV-assisted 6G networks remains an open and exciting future research direction.

\subsubsection{Large Language Models (LLMs)}
Recently, large language models (LLMs) have been shown to be capable of performing a variety of planning tasks. For example, in \cite{Kwon2024IRAL}, the authors show how to use GPT-4 to perform several manipulation tasks with a robotic arm, while \cite{meng-etal-2024-llm} explores the possibility of using LLM in conjunction with the classical A* path planning algorithm to improve performance. LLMs also have the potential to aid in the optimization and coordination in multi-agent systems \cite{Li2024Vicinagearth}. This paradigm presents an intriguing new frontier in the optimization of the deployment of UVs in next-generation communication systems.

\subsubsection{Semantic Communication}
Semantic Communications for UAVs  is an emerging research area that seeks to optimize information exchange between UAVs and other entities by focusing on the meaning or significance of the transmitted data rather than the raw data itself \cite{Luo2022WC}. Unlike traditional communication systems that prioritize bit-level accuracy, semantic communications aim to ensure that the transmitted information is contextually relevant and mission-critical, enabling efficient and effective decision-making in UAV operations. There are many open research questions related to the implementation of semantic communications for UAVs \cite{Xu2024WC} and to specific applications of this new paradigm \cite{Basnayaka2023CAMAD}.

\subsection{Security}
\textcolor{black}{The affordability of hardware and the high probability of LOS communication links make UVs appealing for wireless communications, yet they can also present vulnerabilities. This becomes particularly crucial in military applications, where UVs often carry sensitive information, making privacy and security paramount requirements of system design.}

\textcolor{black}{One challenge lies in the low-cost hardware mounted on UVs, which may struggle to run sophisticated encryption algorithms necessary for ensuring information security. As a result, UVs become susceptible to cybersecurity attacks. These threats may seek to disrupt access to critical services, as in Denial-of-Service (DoS) \cite{vasconcelos2019evaluation} and battery attacks \cite{desnitsky2021simulation}, or they may compromise the privacy of sensitive information, as in spoofing attacks \cite{he2016communication}.}

However, even if computational resources are available, traditional cryptographic approaches may be difficult to implement on a UV. This is because they rely on secret keys, which are difficult to distribute in a wireless network.
Physical Layer Security (PLS) approaches may thus be more applicable in such scenarios. 
By exploiting the physical characteristics of the wireless channel, PLS design aims to enable the legitimate destination to obtain the source information successfully, while preventing an eavesdropper (ED), or unauthorized user from decoding the information \cite{wyner1975wire}. PLS system design aims to maximize the secrecy rate, which is the rate the legitimate receiver receives minus the rate that the eavesdropper receives, subject to certain system constraints.
PLS design of  communication systems has been well investigated \cite{Fakoorian2011solutions,Dong2010improving, Zheng2011optimal,Li2011oncooperative, Zheng2013improving, Khisti2010secure, Goel2008guaranteeing}. One approach to ensure PLS is \textit{cooperative jamming}, where trusted relays act as helpers and beamform artificial noise (AN), aiming to degrade the ED's channel \cite{Fakoorian2011solutions,Dong2010improving,Zheng2011optimal,Li2011oncooperative,Zheng2013improving}. Another approach is for the source to beamform AN along with the intended information for users, in a way that the users do not experience interference \cite{Khisti2010secure,Goel2008guaranteeing}.
PLS design for (Dual Function Radar and Communication (DFRC) systems has been considered in \cite{Su2021secure,Su2022secure}, while the design of DFRC aided by an RIS is investigated in \cite{li2023irs}.

Encryption and  PLS focus on ensuring that information is not intercepted by malicious receivers. However, in military environments, UVs face an additional requirement: covertness. Covertness extends security efforts from merely preventing the interception and decoding of sensitive information to ensuring that the transmission itself goes undetected \cite{chen2023covert}. Historically, many Low Probability of Detection/Low Probability of Interception (LPD/LPI) techniques have been developed for covert communications. However, the mobility of UVs and the characteristics of LOS channels pose additional challenges for maintaining covertness in UV wireless communications. On the positive side, the controllable nature of UVs motion can be leveraged and designed to enhance covertness. The work in \cite{9382022} provides two different paradigms for covertness in UV-assisted networking: The first method is an iterative analytical approach that optimizes the UV motion, transmit power and time slot partition in order to maximize the covert rate. The second approach employs a jammer and optimizes its jamming power, transmit power, and location for covertness. Other works consider covert communication with UV-mounted RIS~\cite{10271264}, with uncertainty around the warden's location~\cite{Li2022ICC}, or with a game-theoretical framing~\cite{Du2022ICC}.

\textbf{Takeaways:} This section explores key challenges and opportunities for deploying UVs in 6G systems, focusing on modeling, computation, and security. As 6G networks demand advanced functionalities, UV deployment requires updates to current models, computational techniques, and security frameworks to meet new requirements. Modeling advancements are crucial for capturing high-frequency channel behavior, detailed UV kinematics, and integrating digital twins for more realistic simulations. Computational challenges include addressing limited UV resources, real-time optimization for large UV networks, and privacy concerns. Solutions involve edge computing, federated learning, and leveraging  LLMs for multi-agent coordination. Security emphasizes the need for lightweight encryption,  PLS, and covert communication techniques to mitigate vulnerabilities, particularly in sensitive contexts such as military applications. Together, these challenges highlight exciting research opportunities to optimize UV operations in next-generation networks.


\section{Conclusions}\label{sec:conclusions}

The tech world eagerly anticipates the advent of the 6th Generation of wireless networking, promising transformative capabilities and applications. The success of this revolution, from closing of the digital divide to fueling the industrial sector, hinges on diverse technological advancements. Unmanned  Vehicles emerge as a pivotal technology, especially crucial for extended coverage and connectivity. While UVs enhance wireless infrastructure, they also introduce novel challenges requiring attention from academia and industry.

This paper offers a comprehensive view of settings and scenarios at the intersection of UVs and wireless networking, accompanied by an in-depth exploration of corresponding problem formulations and methods. To present this information effectively, the central thrust of the paper has been the unification of the problem space, providing a cohesive framework to understand the use cases, problem formulations, and necessary tools. Overall, the paper can equip researchers, engineers, and stakeholders with a clear understanding of both the opportunities and challenges inherent in the convergence of unmanned vehicles and 6G networks, paving the way towards a successful integration.

\bibliographystyle{IEEEtran}
\bibliography{main}

\end{document}